\documentclass[english,aps,prb,citeautoscript,preprint]{revtex4-1}
\usepackage[T1]{fontenc}
\usepackage[latin9]{inputenc}
\usepackage[active]{srcltx}
\usepackage{color}
\usepackage{units}
\usepackage{amsmath}
\usepackage{graphicx}
\usepackage{subscript}

\makeatletter

\providecommand{\tabularnewline}{\\}

\setcitestyle{super}
\usepackage{amsmath}
\usepackage{babel}

\@ifundefined{showcaptionsetup}{}{%
 \PassOptionsToPackage{caption=false}{subfig}}
\usepackage{subfig}
\makeatother

\usepackage{babel}
\begin{document}

\title{{\normalsize{}Theory of Linear Optical Absorption in Diamond Shaped
Graphene Quantum Dots}}

\author{{\normalsize{}Tista Basak{*}, Himanshu Chakraborty and Alok Shukla}}

\affiliation{Department of Physics, Indian Institute of Technology, Bombay, Mumbai-400076,
INDIA.}

\altaffiliation{Mukesh Patel School of Technology Management and Engineering, NMIMS University, Mumbai-56, India; tista.basak@nmims.edu, shukla@phy.iitb.ac.in, chakraborty.himanshu@gmail.com}

\begin{abstract}
{\normalsize{}In this paper, optical and electronic properties of
diamond shaped graphene quantum dots (DQDs) have been studied by employing
large-scale electron-correlated calculations. The computations have
been performed using the $\pi$-electron Pariser-Parr-Pople model
Hamiltonian, which incorporates long-range Coulomb interactions. The
influence of electron-correlation effects on the ground and excited
states has been included by means of the configuration-interaction
approach, used at various levels. Our calculations have revealed that
the absorption spectra are red-shifted with the increasing sizes of
quantum dots. It has been observed that the first peak of the linear
optical absorption, which represents the optical gap, is not the most
intense peak. This result is in excellent agreement with the experimental
data, but in stark contrast to the predictions of the tight-binding
model, according to which the first peak is the most intense peak,
pointing to the importance of electron-correlation effects. }{\normalsize \par}
\end{abstract}


\maketitle

\section{{\normalsize{}Introduction}}

Graphene, a two-dimensional, one atom thick layer of graphite, with
carbon atoms arranged in a honeycomb lattice, has attracted enormous
attention among researchers in recent years due to the possibilities
of appealing applications in the field of nanoelectronics\cite{Wessely201483}.
However, a major drawback of pure graphene from the viewpoint of electronic
devices, in general, and opto-electronic devices, in particular, is
its zero band gap. This problem has stimulated tremendous amount of
experimental and theoretical efforts in attempting to formulate techniques
to introduce a band gap in graphene\cite{RevModPhys.81.109Castro.Neto}.
It has been observed that reducing the dimensionality of graphene
opens up the band gap on account of quantum confinement. One-dimensional
periodic graphene nanostructures such as graphene nanoribbons (GNRs)
have band gaps ranging from zero (metallic) to rather large values
(semiconducting), depending on the width of the ribbon, and the nature
of edge termination\cite{PhysRevB.73.045432Ezawa.2006}. Opening up
the band-gap further, by reducing the dimension of GNRs, has been
made possible by fabrication of stable zero-dimensional graphene quantum
dots (GQDs)\cite{Lifabricationdoi:10.1021/jz100862f,PPSC:PPSC201300252Bacon,HanjunSun2013433}.
The band-gaps of GNRs are $\approx$0.4 eV, while the band-gaps of
GQDs can be tuned up to $\approx$3 eV by decreasing their size \cite{doi:10.1021/nn302878rSungKim}.
This appealing feature of GQDs enhances the prospects of utilization
of such materials in lasers, light-emitting diodes (LEDs), solar cells,
bio-imaging sensors,\cite{C1CC11122ATang[8]} and optically addressable
qubits in quantum information science.\cite{C2CC00110AShen} Since,
electronic excitation determine the photophysics of GQDs which is
vital for all these applications, it is essential to have a detailed
understanding of their low-lying excited states which can be probed
by optical means.

Because of aforesaid possibilities of applications of GQDs, significant
studies, experimental, as well as theoretical, on the electronic and
optical properties of GQDs have been performed lately. Experimental
studies on GQDs (in the size range of 5-35 nm) by Kim \emph{et al}.\cite{doi:10.1021/nn302878rSungKim}
have revealed that while the absorption peak energies decrease with
increasing size of the GQDs, the photoluminescence (PL) spectra exhibit
a decrease in energy as the average size of GQDs increases up to \ensuremath{\sim}17
nm, followed by an increase in the peak energy with increasing average
size of GQDs. They accredited this abnormal behavior of PL spectra
to the presence of edge variations associated with size-dependent
shape in GQDs. However, single-particle spectroscopic measurements
carried out by Xu \emph{et al}.\cite{doi:10.1021/nn4053342QinfengXu}
have shown that size differences of GQDs do not affect the peak positions
and spectral line-shapes. Experiments on photoexcited GQDs have also
indicated that emission intensity decreases when GQDs are excited
to singlet states ($S_{1}$, $S_{2}$, $S_{3}$), while it increases
sharply when they are excited to $S_{4}$, or higher excited states
\cite{doi:10.1021/nl101474dMuellerNanoletters}. In addition, several
experimental studies have shown that the optical band-gap is dependent
on the size of the GQDs, giving rise to different excitation/emission
spectra as well as PL spectra of different colors. \cite{Kwon-band-gap,Li-band-gap}
Further, it has been observed that PL behavior is strongly associated
with the presence of microstructures in GQDs and hence, are affected
by edge effects as well as emission sites \cite{ChenChemCommun}.
Thus, it is essential to have a detailed knowledge of the atoms which
significantly contribute to the optical band-gap. 

As far as theoretical studies are concerned, Yamijala \emph{et al}.\cite{doi:10.1021/jp406344zSwapanPati}
have performed a detailed study of the structural stability, electronic,
magnetic and optical properties of rectangular shaped graphene quantum
dots as a function of their size, as well as under the application
of electric field, using first-principles density functional theory
(DFT). However, DFT based calculations are known to underestimate
the band gap, and provide a reasonable description of the excited
states only when they do not exhibit significant configuration mixing.
Yan \emph{et al}.\cite{doi:10.1021/jz200450rXinYan} have employed
a tight-binding (TB) model to predict the band gap of GQDs as a function
of their size. Theoretical calculations using the TB model have also
been utilized to study the optical properties of hexagonal \cite{PhysRevB.77.235411Zhang}
and triangular graphene quantum dots as a function of their size and
type of edge \cite{PhysRevB.74.121409Yamamoto}. However, TB method
is unreliable in predicting low-lying excited states because it does
not include electron-electron interactions. In addition, calculations
of optical properties of large graphene quantum dots employing first-principles
DFT have also been performed by Schumacher. \cite{PhysRevB.83.081417StefanSchumacher}
Recent first principles DFT calculations on diamond-shaped graphene
nanopatches have revealed that these systems display well-defined
magnetic states which can be selectively tuned by the application
of electric field.\cite{PhysRevB.82.201411} However, to the best
of our knowledge, till date there is no existing literature (experimental
as well as theoretical) on the optical properties of diamond shaped
graphene quantum dots. 

Motivated by aforementioned theoretical and experimental studies of
GQDs, in this work we present a systematic study of the electronic
structure and the optical properties of diamond-shaped graphene quantum
dots (DQDs) which exhibit a mixture of zigzag edges and armchair corners,
and we hope that our studies will motivate experimentalists to explore
optical properties of these nanostructures. For this purpose, we have
employed a methodology based upon Pariser-Parr-Pople (PPP) model Hamiltonian,\cite{ppp-pople,ppp-pariser-parr}
which is an effective $\pi$-electron model, including long-range
electron-electron interactions. We have used this approach in several
works in our group dealing with conjugated polymers,\cite{PhysRevB.65.125204Shukla65,PhysRevB.69.165218Shukla69,PhysRevB.71.165204Priya_t0,:/content/aip/journal/jcp/131/1/10.1063/1.3159670Priyaanthracene,doi:10.1021/jp408535u,himanshu-triplet,sony-acene-lo}
polyaromatic hydrocarbons,\cite{doi:10.1021/jp410793rAryanpour,:/content/aip/journal/jcp/140/10/10.1063/1.4867363Aryanpour}
graphene nanoribbons,\cite{PhysRevB.83.075413Kondayya,PhysRevB.84.075442Gundra}
and graphene nanodisks.\cite{Sony2010821} PPP model has an advantage
over the TB model in that it incorporates long-range Coulomb interactions
among the $\pi$-electrons, essential for taking into account influence
of electron correlation effects. Further, it considers the interactions
of $\pi$-electrons with a minimal basis, therefore, as compared to
\emph{ab initio} approaches, it yields highly accurate results with
fewer computational resources. In this work, we present theoretical
calculations of the electronic structure and linear optical absorption
spectra of DQDs of varying sizes employing a configuration-interaction
(CI) methodology,\cite{PhysRevB.65.125204Shukla65,PhysRevB.69.165218Shukla69,PhysRevB.71.165204Priya_t0,:/content/aip/journal/jcp/131/1/10.1063/1.3159670Priyaanthracene,doi:10.1021/jp408535u,himanshu-triplet,sony-acene-lo}
so as to account for electron-correlation effects in their ground
and excited states. As far as experiments are concerned, it is impossible
to synthesize bare graphene quantum dots of high symmetry, because\textcolor{black}{,}\textcolor{red}{{}
}due to the dangling bonds, edges will undergo significant reconstruction,
leading to distorted shapes. Nevertheless, several polycyclic aromatic
hydrocarbons (PAHs) have been synthesized which are nothing but graphene
quantum dots of high symmetry, but with edges passivated by hydrogen
atoms,\cite{pah-database-malloci} a few of which we had studied in
earlier\textcolor{red}{{} }works.\cite{doi:10.1021/jp410793rAryanpour,:/content/aip/journal/jcp/140/10/10.1063/1.4867363Aryanpour}
Of the quantum dots considered here, hydrogen passivated counterpart
of DQD with 16 carbon atoms (DQD-16, henceforth) is called pyrene,
while that of DQD with 30 carbon atoms (DQD-30) is known as dibenzo{[}bc,kl{]}coronene,
both of which have been well-studied in the chemical literature.\cite{pah-database-malloci}
A large number of experimental measurements of optical absorption
of pyrene in vapor,\cite{Thony} solution,\cite{pyrene-exp-indrasen,pyrene-photocad,BasuRay2006248,Ram20092252}
and matrix isolated phases\cite{Salama_1993,Gudipati_1993,pyrene-exp-vala,Halasinski_2005}
have been performed, and our results on DQD-16 are in excellent agreement
with them.\textcolor{red}{{} }Clar and Schmidt\cite{clar-dbcoronene}
measured the gas-phase absorption spectrum of dibenzo{[}bc,kl{]}coronene,
and our calculations on DQD-30 are in very good agreement with the
experimental results. We also computed the absorption spectrum of
next larger quantum dot DQD-48, whose structural properties have been
studied theoretically by several authors,\cite{New_Paper_Boersma_C48H18,New_Paper_C48H18_Deepak_H-L,New_paper_C48H18_Denis,New_Paper_C48H18_Pathak2007898}
 but it has not been synthesized as yet. 

Theoretically, Canuto \emph{et al.,}\cite{Canuto} computed the absorption
spectrum of pyrene employing the intermediate neglect of differential
overlap (INDO/S) semi-empirical quantum mechanical technique along
with singles configuration interaction (SCI) method, while Gudipati
\emph{et al.,}\cite{Gudipati_1993} calculated the excitation energies
and oscillator strengths of pyrene using the complete neglect of differential
overlap (CNDO/S) model, coupled with the truncated singles and doubles
configuration interaction (SDCI) method. Parac \emph{et al.,}\cite{Parac200311}
and Malloci \emph{et al.,}\cite{pyrene-theory-malloci} employed the
time dependent density functional theory (TDDFT) technique to compute
the excitation energies and photoabsorption cross-sections of pyrene,
respectively. Malloci \emph{et al.,}\cite{dibenzocoronene-database}
also calculated the photoabsorption cross-section of dibenzo{[}bc,kl{]}coronene
using the TDDFT technique. Several authors have studied the structural
stability of the PAH equivalent of DQD-48 (C$_{48}$H$_{18}$) by
employing first-principles DFT-based methodologies.\cite{New_Paper_Boersma_C48H18,New_Paper_C48H18_Deepak_H-L,New_paper_C48H18_Denis,New_Paper_C48H18_Pathak2007898}
Additionally, Karki \emph{et al}.\cite{New_Paper_C48H18_Deepak_H-L}
also studied the variation of the optical gap with increasing size
of the PAH clusters, while Boersma \emph{et al}.,\cite{New_Paper_Boersma_C48H18}
and Pathak \emph{et al}.\cite{New_Paper_C48H18_Pathak2007898} calculated
their infrared spectra. Denis\emph{ et al}.,\cite{New_paper_C48H18_Denis}
analysed the effect of addition of azomethine ylide on the binding
energy of $\mbox{C}$$_{48}\mbox{H}_{18}$. 

Based upon our calculations, we predict the variation in the behavior
of linear absorption spectrum with increasing size of DQDs, and our
results are in significant variance with the predictions of the TB
model. We also identify the atoms which play a significant role in
the band gap, and thus the optical spectrum of the DQDs. 

The remainder of this paper is organized as follows. In section \ref{sec:theory},
we present a brief overview of the theoretical methodology adopted
by us. In section \ref{sec:results}, we present and discuss the results,
followed by conclusions in section \ref{sec:conclusions}. An Appendix
representing detailed information about many-particle wave-functions
of excited states contributing to the optical absorption peaks is
presented at the end of the paper.

\section{{\normalsize{}Computational Details}}

\label{sec:theory}

\begin{figure}
\subfloat{\includegraphics[width=5cm]{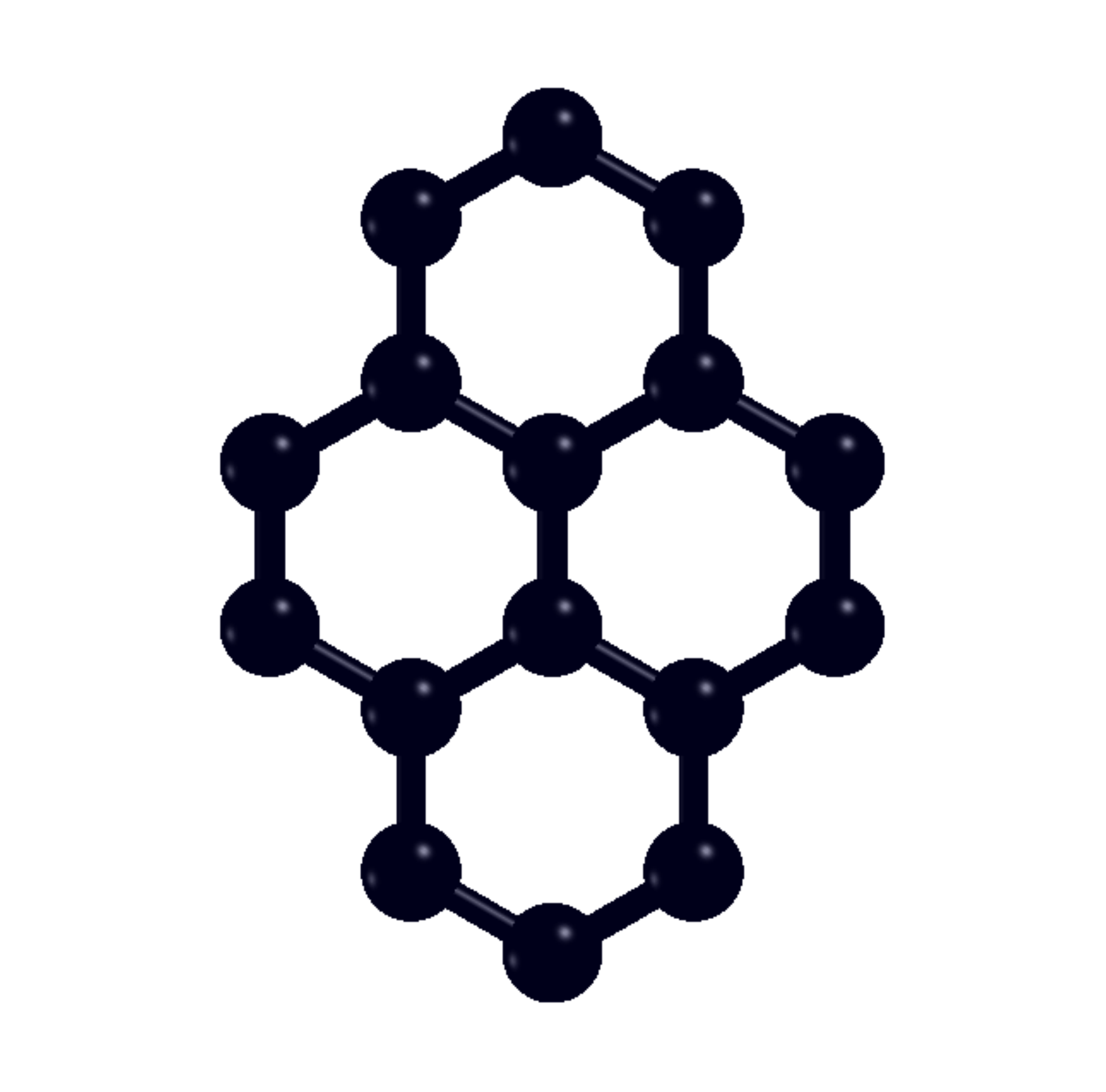}

}\subfloat{\includegraphics[width=7cm]{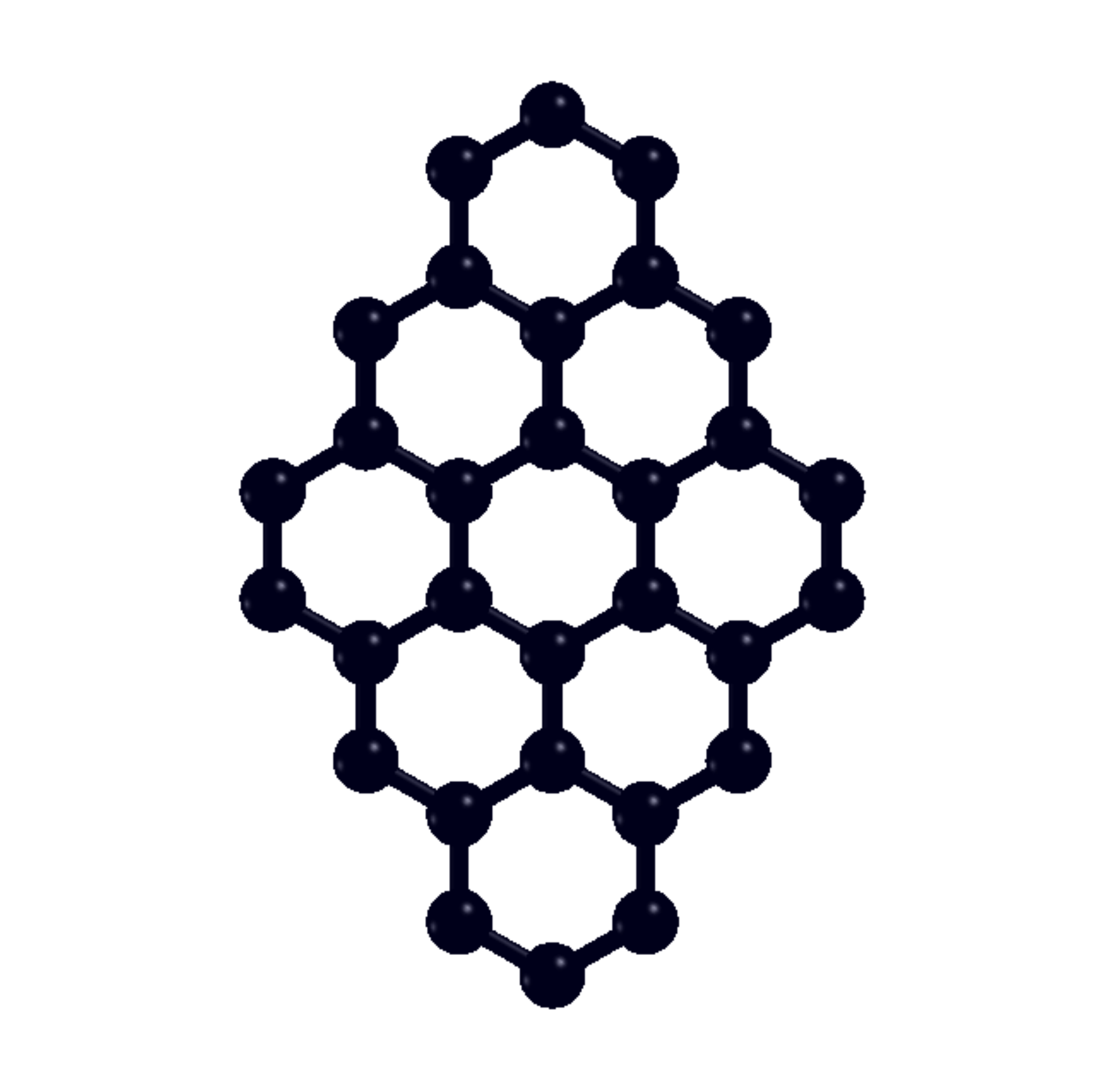}

}\protect\subfloat{

\includegraphics[width=8cm]{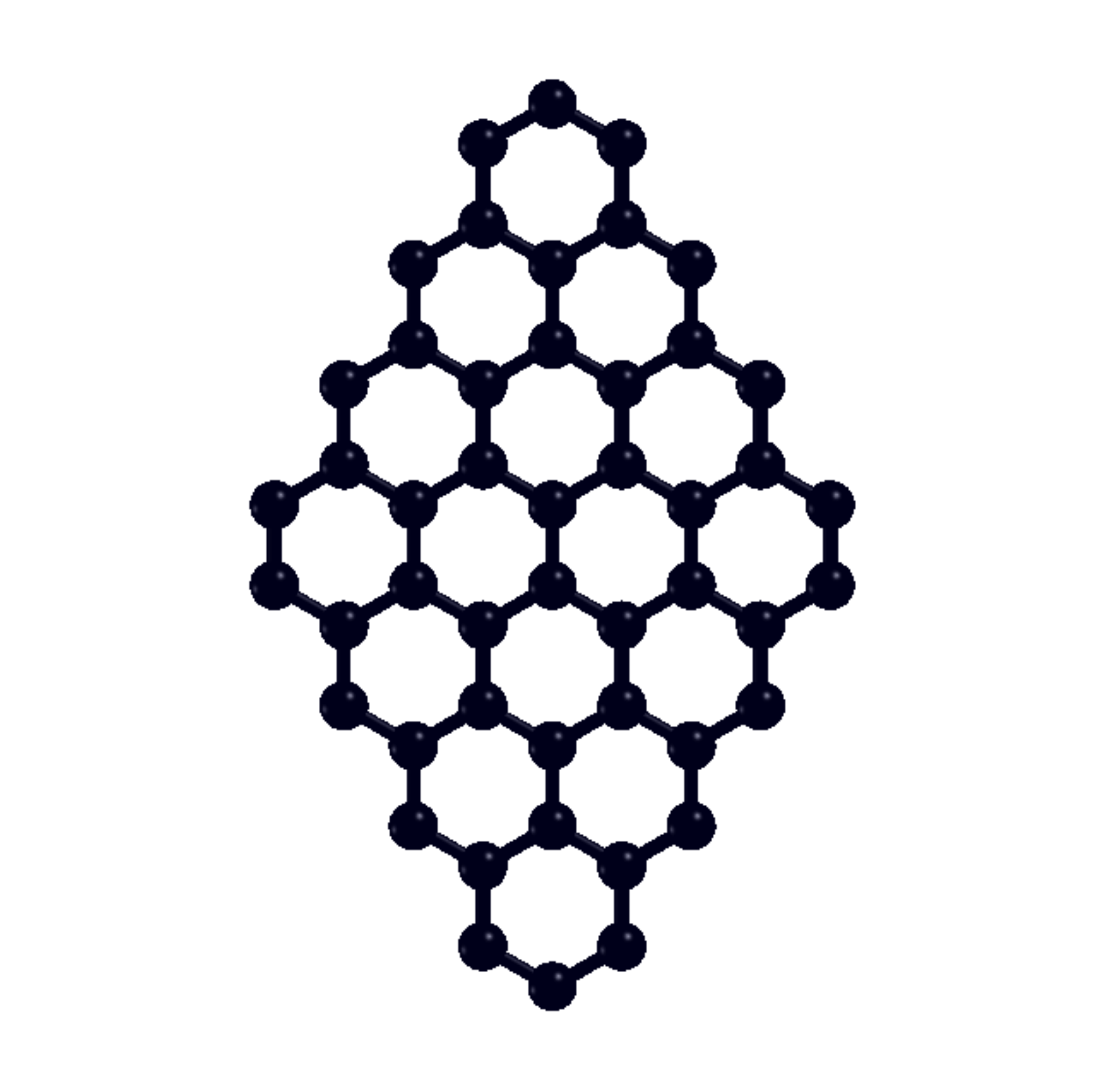}

}\caption{Schematic \label{fig:Schematic-diagram-of}diagram of diamond shaped
graphene quantum dots consisting of 16 atoms, 30 atoms and 48 atoms,
respectively.}
\end{figure}

The schematic diagram of the geometry of DQDs considered in this work
is given in Fig.\ref{fig:Schematic-diagram-of}. Different DQDs can
be identified by the total number of carbon atoms $n$, and will be
denoted as DQD-$n$, henceforth. In our calculations, all quantum
dots are assumed to lie in the $x-y$ plane, with the shorter diagonal
of the DQD assumed to be along the $x$-axis, and the longer one along
the $y$-axis. All carbon-carbon bond lengths and bond angles have
been chosen as 1.4 \AA, and 120\textsuperscript{o}, respectively.
The point group of DQDs is $D$$_{2h}$, with $1{}^{1}$A$_{g}$ being
the ground state. Then, as per electric-dipole selection rules, the
symmetries of the one-photon excited states are $^{1}$B$_{2u}$ and
$^{1}$B$_{3u}$. 

These calculations have been carried out by employing the PPP model
Hamiltonian\cite{ppp-pople,ppp-pariser-parr}, given by

\begin{equation}
\mbox{\mbox{\mbox{\mbox{\ensuremath{H}\ensuremath{=}\ensuremath{-}\ensuremath{\sum_{i,j,\sigma}t_{ij}\left(c_{i\sigma}^{\dagger}c_{j\sigma}+c_{j\sigma}^{\dagger}c_{i\sigma}\right)}\ensuremath{+}\ensuremath{U\sum_{i}n_{i\uparrow}n_{i\downarrow}}\ensuremath{+}\ensuremath{\sum_{i<j}V_{ij}(n_{i}-1)(n_{j}-1)}}}}}\label{eq:ppp}
\end{equation}

where $c_{i\sigma}^{\dagger}($c$_{i\sigma})$ creates (annihilates)
a $\pi$ orbital of spin $\sigma$, localized on the \emph{i}th carbon
atom while the total number of electrons with spin $\sigma$ on atom
$i$ is indicated by $n$$_{i}=\sum_{\sigma}c_{i\sigma}^{\dagger}c_{i\sigma}$.
The second and third terms in Eq. \ref{eq:ppp} denote the electron-electron
repulsion terms, with the parameters $U$, and $V_{ij}$ representing
the on-site, and the long-range Coulomb interactions, respectively.
The matrix elements $t_{ij}$ depicts one-electron hops, which in
our calculations, have been restricted to nearest neighbours, with
the value $t{}_{0}$= 2.4 eV, consistent with our earlier calculations
on conjugated polymers,\cite{PhysRevB.65.125204Shukla65,PhysRevB.69.165218Shukla69,PhysRevB.71.165204Priya_t0,:/content/aip/journal/jcp/131/1/10.1063/1.3159670Priyaanthracene,doi:10.1021/jp408535u,himanshu-triplet,sony-acene-lo}
and polyaromatic hydrocarbons.\cite{doi:10.1021/jp410793rAryanpour,:/content/aip/journal/jcp/140/10/10.1063/1.4867363Aryanpour}

Parameterization of the Coulomb interactions is done according to
the Ohno relationship \cite{Theor.chim.act.2Ohno}

\begin{equation}
V_{ij}=U/\kappa_{i,j}(1+0.6117R_{i,j}^{2})^{\nicefrac{1}{2}}
\end{equation}

where $\kappa_{i,j}$ represents the dielectric constant of the system
which replicates screening effects, $U$ as described above is the
on-site electron-electron repulsion term, and $R_{i,j}$ is the distance
(in \AA) between the $i$th and $j$th carbon atoms. In the present
work, we have done calculations adopting both ``screened parameters''\cite{PhysRevB.55.1497Chandross}
with $U=8.0$ eV, $\kappa_{i,j}=2.0(i\neq j)$ and $\kappa_{i,i}=1.0$,
as also the ``standard parameters'' with $U=11.13$ eV, and $\kappa_{i,j}=1.0$.
We observe that our calculations employing the screened parameters,
proposed by Chandross and Mazumdar,\cite{PhysRevB.55.1497Chandross}
are in better agreement with the experimental results, as compared
to those performed using standard parameters, consistent with the
trends observed in our earlier works as well.\cite{PhysRevB.69.165218Shukla69,doi:10.1021/jp408535u} 

The first step of our calculations is to find the self-consistent
solutions at the Restricted Hartree-Fock (RHF) level, employing the
PPP Hamiltonian (\emph{cf}. Eq. \ref{eq:ppp}), using a code developed
in our group.\cite{Sony2010821} These solutions, in which electrons
occupy the lowest energy orbitals, comprise the HF ground state. This
is followed by correlated calculations at the quadruple configuration
interaction (QCI) level, or at the multi-reference singles-doubles
configuration interaction (MRSDCI) level, depending upon the size
of DQD. In the QCI approach, up to quadruple excitations from the
HF ground state are considered, and, thus, it requires significant
amount of computational resources. Therefore, QCI calculations can
be performed only for small systems (in our case, for DQD-16). For
larger DQDs, MRSDCI approach has been employed. In MRSDCI calculations,
singly and doubly excited configurations from the reference configurations
of the selected symmetry subspace are considered while generating
the CI matrix.\cite{peyerimhoff_CI,peyerimhoff_energy_CI} Subsequently
these CI wave functions are used to compute transition electric dipole
matrix elements between various states, required for computing the
optical absorption spectra. From the calculated spectra, important
excited states giving rise to various peaks are identified, and the
dominant reference configurations contributing to these excited states
are included to enhance the new reference space. This procedure is
iterated until the desired absorption spectrum converges to an acceptable
tolerance. With the increasing sizes of the DQDs, the number of molecular
orbitals of the DQD increases, leading to an increase in the size
of the CI expansion. Therefore, to make calculations feasible, the
frozen orbital approximation was adopted for DQD-48, with the lowest
two occupied orbitals frozen, and highest two virtual orbitals deleted,
so as to retain the particle-hole symmetry. 

The formula employed for the calculation of the ground state optical
absorption cross-section $\sigma(\omega)$, assumes a Lorentzian line
shape

\begin{equation}
\sigma(\omega)=4\pi\alpha\underset{i}{\sum}\frac{\omega_{i0}\left|\left\langle i\left|\mathbf{\hat{e}.r}\right|0\right\rangle \right|^{2}\gamma}{\left(\omega_{i0}-\omega\right)^{2}+\gamma^{2}},
\end{equation}

where $\omega$ denotes incident radiation frequency, $\hat{{\bf e}}$
denotes its polarization direction, ${\bf r}$ is the position operator,
$\alpha$ is the fine structure constant, $0$ and $i$ denote, respectively,
the ground and the excited states, $\omega_{i0}$ is the frequency
difference between those states, and $\gamma$ is the absorption line-width.

\section{{\normalsize{}Results And Discussion}}

\label{sec:results}

In this section, we present the results obtained from CI calculations
for DQDs of varying sizes, ranging from DQD-16 to DQD-48. In order
to acquaint the reader with the precision of our MRSDCI or QCI calculations,
the sizes of the resultant CI matrix for different symmetries for
the DQDs considered here are given in Table \ref{tab:Total-size-of}.
QCI method was employed for carrying out calculations on $^{1}A_{g}$
and $^{1}B_{2u}$ manifolds of DQD-16, while rest of computations
on DQD-16, DQD-30, and DQD-48, were carried out by adopting MRSDCI
methodology. Sizes of the CI matrix indicate that electron-correlation
effects were well accounted for in these calculations. 

\begin{center}
\begin{table}
\centering{}\protect\caption{Dimensions of CI matrices for DQDs of varying sizes, for various symmetry
manifolds. QCI method was employed for carrying out calculations on
$^{1}A_{g}$ and $^{1}B_{2u}$ manifolds of DQD-16, while for rest
of the calculations MRSDCI method was used.}

\begin{tabular}{|c|c|c|c|}
\hline 
Number of atoms in DQD  & \multicolumn{3}{c|}{Dimension of CI matrix}\tabularnewline
\cline{2-4} 
 & $^{1}A_{g}$  & $^{1}B_{2u}$  & $^{1}B_{3u}$ \tabularnewline
\hline 
16  & 73857  & 126279  & 142992 \tabularnewline
\hline 
30  & 215919  & 1564554  & 1359014 \tabularnewline
\hline 
48  & 237030  & 5442399  & 4269236 \tabularnewline
\hline 
\end{tabular}\label{tab:Total-size-of}
\end{table}

\par\end{center}

\subsection{Charge density distribution and optical gap}

In this section we examine the evolution of the band gap, orbital
energy levels, and the charge densities with the size of the DQD.

\subsubsection{Charge Density}

\begin{figure}
\includegraphics[scale=0.7]{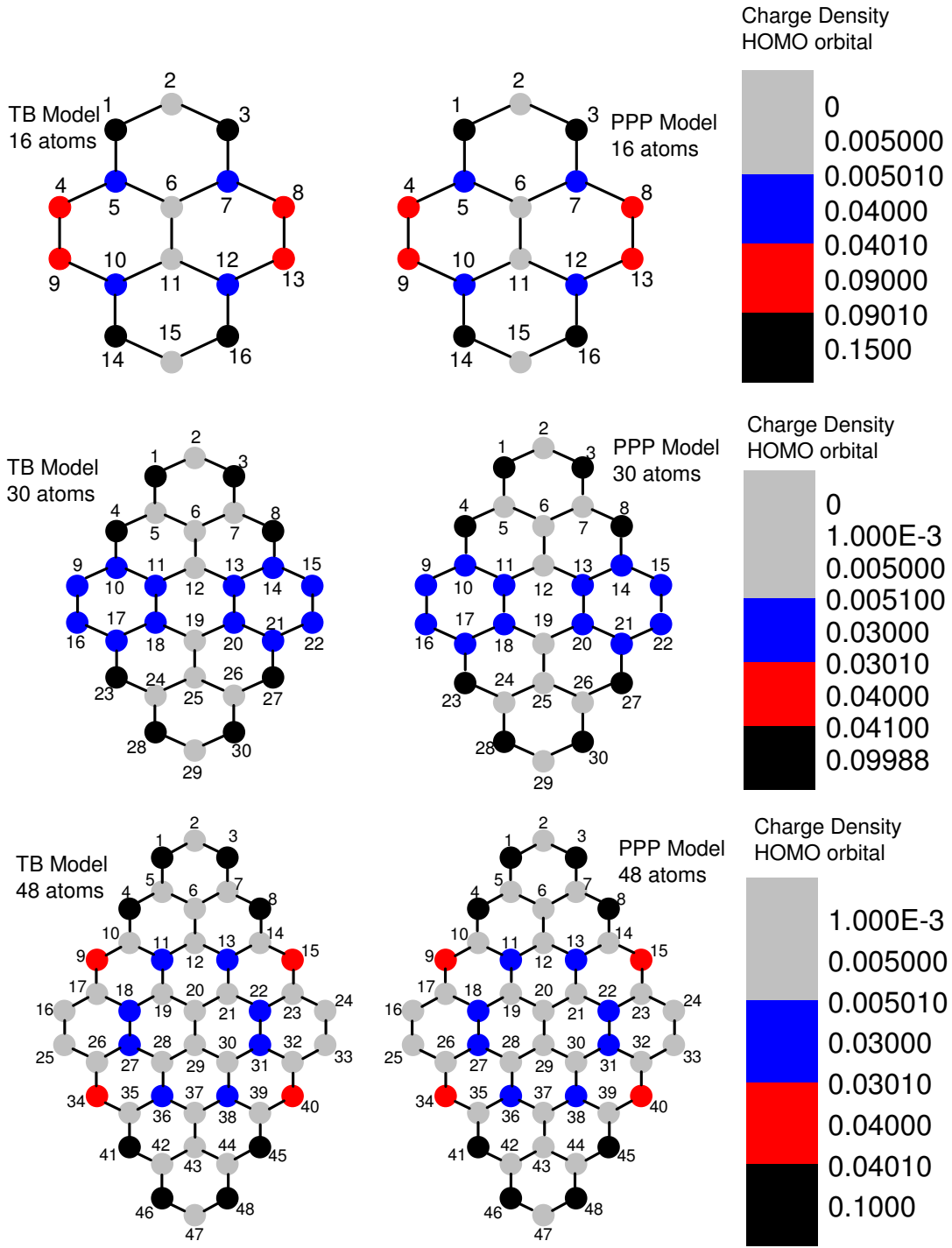}\protect\caption{\label{fig:Charge-density-plot}Charge density bubble plots of HOMO
orbital obtained by employing TB and PPP models for DQDs of varying
sizes. The numbering scheme of atoms for the different DQDs considered
is also presented.}
\end{figure}

The charge density bubble-plots for the HOMO orbital obtained by employing
the TB model and the PPP model for DQDs of varying sizes are presented
in Fig. \ref{fig:Charge-density-plot}. The charge density of the
LUMO orbital is same as that of the HOMO orbital because of the electron-hole
symmetry, and hence has not been shown. The numbering scheme of atoms
for the different DQDs considered is also presented.

It is observed that in case of the smallest quantum dot DQD-16, the
contribution from atomic sites 1, 3, 14 and 16 to the charge density
of HOMO orbital is maximum. These atoms are at the projected corners
of a purely zigzag edge as is evident from \textcolor{red}{Fig. \ref{fig:Charge-density-plot}}.
The charge density contribution from the atomic sites 4, 8, 9, and
13, which are also at the edge of the diamond quantum dot, is less
as compared to that of the atoms mentioned earlier. This can be attributed
to the fact that these atoms give rise to an edge which exhibits both
zigzag as well as armchair nature. With the increasing size of the
quantum dot, the zigzag characteristic of the edge becomes more conspicuous,
leading to increased contribution of the atoms located on the zigzag
edges. This trend is obvious from the dominant contributions to the
charge densities of the HOMO orbital by atoms 4, 8, 23 and 27 for
DQD-30, and atoms 4, 8, 41 and 45, for DQD-48. From this it is evident
that the HOMO-LUMO band-gap, and thus the optical properties of DQDs,
can be tuned if suitable functional groups are attached to these atoms
on the zigzag edges. This is in agreement with results obtained earlier
for the case of graphene nano-ribbons (GNRs),\cite{Tan_GNR} as well
as triangular nanographenes.\cite{PhysRevB.74.121409Yamamoto} 

\begin{figure}
\includegraphics[width=8cm]{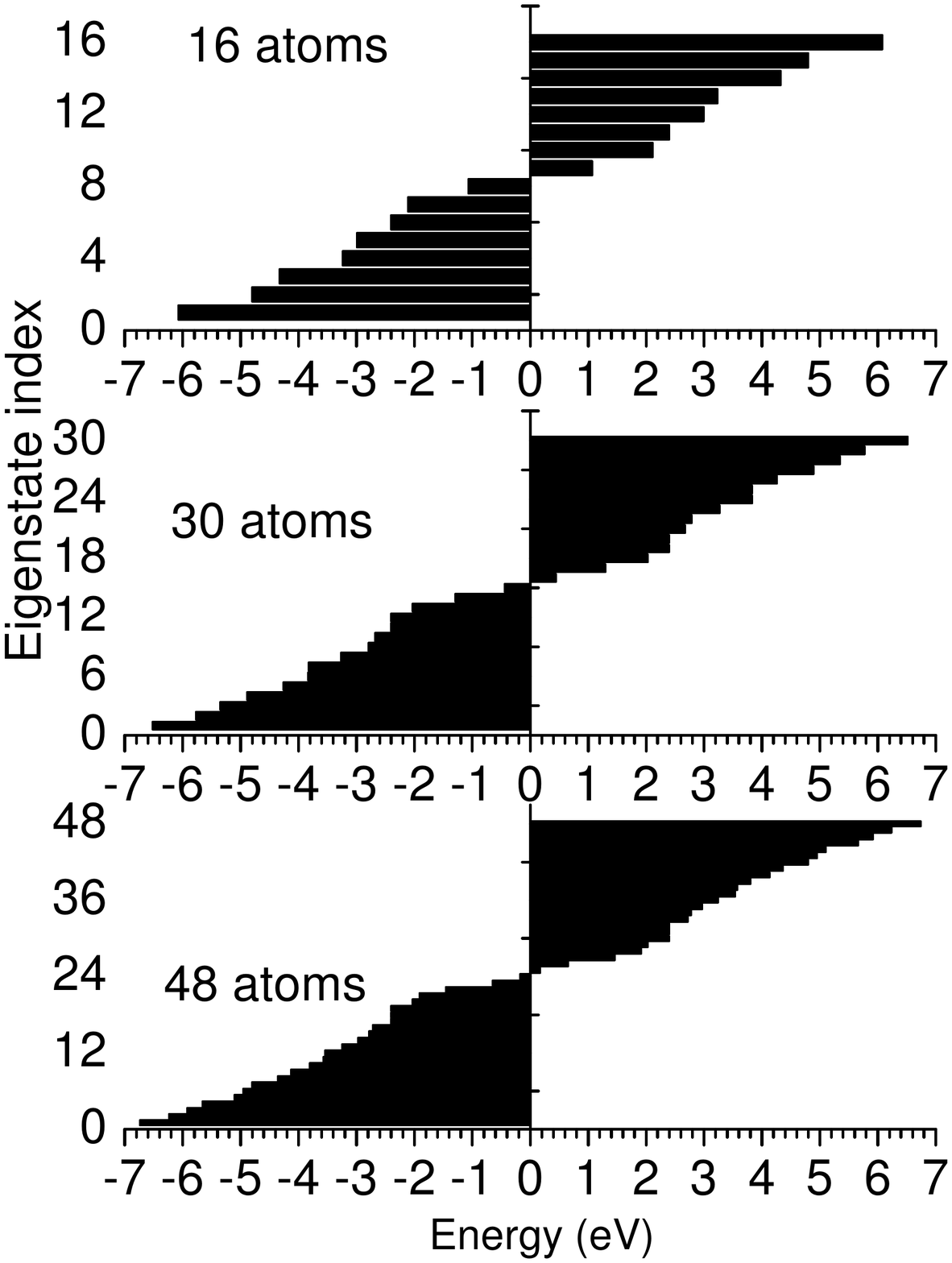}\protect\caption{\label{fig:Energy-level-plots}Energy level plots for DQDs of varying
sizes, calculated using the TB model. }
\end{figure}

Looking at the energy-level diagrams of various DQDs presented in
Fig. \ref{fig:Energy-level-plots}, unlike the case of triangular
nanographenes with zigzag edges, one notices the absence of zero-energy
states.\textcolor{red}{{} }It is also evident from the figure that the
increasing size of DQDs leads to the reduction of the band gap, consistent
with the gapless nature of infinite graphene. 

\subsubsection{Optical gap}

In Table \ref{tab:HOMO-LUMO-band-gap-for}, we present the energy
gap between the highest-occupied molecular orbital (HOMO), and the
lowest-unoccupied molecular orbital (LUMO), for increasing sizes of
DQDs obtained from the TB model, as well as from the PPP model, at
the RHF level. Because the band gap is also the optical gap in DQDs,
in the same table we also present the results of the optical gap of
these systems obtained from our correlated electron CI calculations.
We note the following trends in these results: (a) the gaps decrease
with the increasing DQD size, irrespective of the Hamiltonian or the
method used, (b) gaps obtained from the TB model are significantly
smaller as compared to those obtained from other methods, (c) at the
HF level standard parameter gaps are much larger than those obtained
using screened parameter, (d) optical gaps at the CI level are significantly
redshifted as compared to their HF values, in the standard parameter
calculations. But, for the screened parameter calculations, these
correlation-induced shifts are small, with the CI level gaps of DQD-16
and DQD-30 exhibiting slight redshifts, while that of DQD-48 exhibiting
a small blueshift, and (e) at the CI level the gaps obtained using
the two sets of PPP parameters are in good quantitative agreement
with each other, suggesting the correctness of our correlated electron
approach. This decrease of the gap with the increasing sizes observed
for DQDs is in agreement with previous experimental and theoretical
results obtained for graphene quantum dots of other shapes. \cite{doi:10.1021/jp406344zSwapanPati,PhysRevB.77.235411Zhang} 

\begin{center}
\begin{table}
\centering{}\protect\caption{HOMO-LUMO band gap for increasing size of DQDs obtained from the TB
model and the PPP model. In case of PPP model, the gap is calculated
both at the HF and CI level, using the standard (Std) as well as the
screened (Scr) parameters. At the CI level, the gap is identified
with the optical gap.}

\begin{tabular}{|c|ccccc|}
\hline 
System  & HOMO-LUMO gap (eV) & \multicolumn{2}{c}{HOMO-LUMO gap (eV)} & \multicolumn{2}{c|}{Optical gap (eV)}\tabularnewline
 & (TB model) & \multicolumn{2}{c}{(PPP-HF) } & \multicolumn{2}{c|}{(PPP-CI) }\tabularnewline
\hline 
 &  & Scr & Std & Scr & Std\tabularnewline
\hline 
DQD-16  & 2.14 & 3.91 & 7.26 & 3.60  & 3.74 \tabularnewline
\hline 
DQD-30  & 0.89 & 2.11 & 4.65 & 2.08  & 2.31 \tabularnewline
\hline 
DQD-48  & 0.34 & 1.10 & 2.81 & 1.40  & 1.61 \tabularnewline
\hline 
\end{tabular}\label{tab:HOMO-LUMO-band-gap-for}
\end{table}

\par\end{center}

\subsection{Linear Absorption spectrum}

In this section, we first elucidate the salient features of the linear
optical spectra of DQDs of varying sizes computed within the framework
of the independent-electron TB model and PPP model at the HF level,
which will allow us to gauge the influence of electron-correlation
effects in the PPP model-based CI calculations, presented thereafter.

\subsubsection{Calculations at the tight-binding and HF level}

The absorption spectrum obtained from TB model and PPP model at HF
level employing screened parameters (Fig. \ref{fig:Computed-optical-absorption})
exhibits the following characteristics: 
\begin{enumerate}
\item The absorption spectrum is red-shifted with increase in size of the
DQD, in agreement with quantum confinement effect. This red-shift
is more pronounced at the PPP-HF level as compared to that obtained
by the TB model. In addition, the absorption spectrum at the PPP-HF
level is blue-shifted compared to the one computed using the TB model
because it is well known that the HF theory overestimates energy gaps,
as was also observed in earlier works. \cite{Hawrylak_PhysRevB82,Hawrylak_PhysRevB89}
\item The pattern of the absorption spectra at the PPP-HF level is similar
to that obtained by the TB model. However, the absolute intensities
of the peaks at the PPP-HF level are lesser as compared to those obtained
by the TB model. 
\item The first peak at the TB and PPP-HF level is always $y$-polarized,
and corresponds to excitation of a single electron from the HOMO ($H$)
orbital to the LUMO ($L$) orbital. This peak is also the most intense
peak in the calculated spectra, which is in stark contrast with the
experimental results obtained for pyrene and dibenzo{[}bc,kl{]}coronene.\cite{pyrene-exp-indrasen,pyrene-exp-vala,pyrene-photocad,BasuRay2006248,Salama_1993,dbcoronene-database,clar-dbcoronene,Gudipati_1993,Halasinski_2005,Ram20092252,Thony}
\item With the increasing size of the DQDs, the intensity of the first peak
increases enormously as compared to the other peaks in the absorption
spectrum. For example, in DQD-30/DQD-48, the relative intensity of
the peaks starting from the second one is much smaller as compared
to DQD-16, irrespective of the Hamiltonian employed. 
\item We also note that $x$-polarized peaks are degenerate, while $y$-polarized
peaks exhibit no degeneracy. For example, in case of DQD-16, the second
peak is $x$-polarized and is due to degenerate excitations $|H-1\rightarrow L\rangle$
and $|H\rightarrow L+1\rangle$, while the third peak is $y$-polarized,
and is due to nondegenerate excitation $|H-1\rightarrow L+1\rangle$.
\end{enumerate}
\begin{flushleft}
\begin{figure}
\vspace{-5cm}

\includegraphics[scale=0.75]{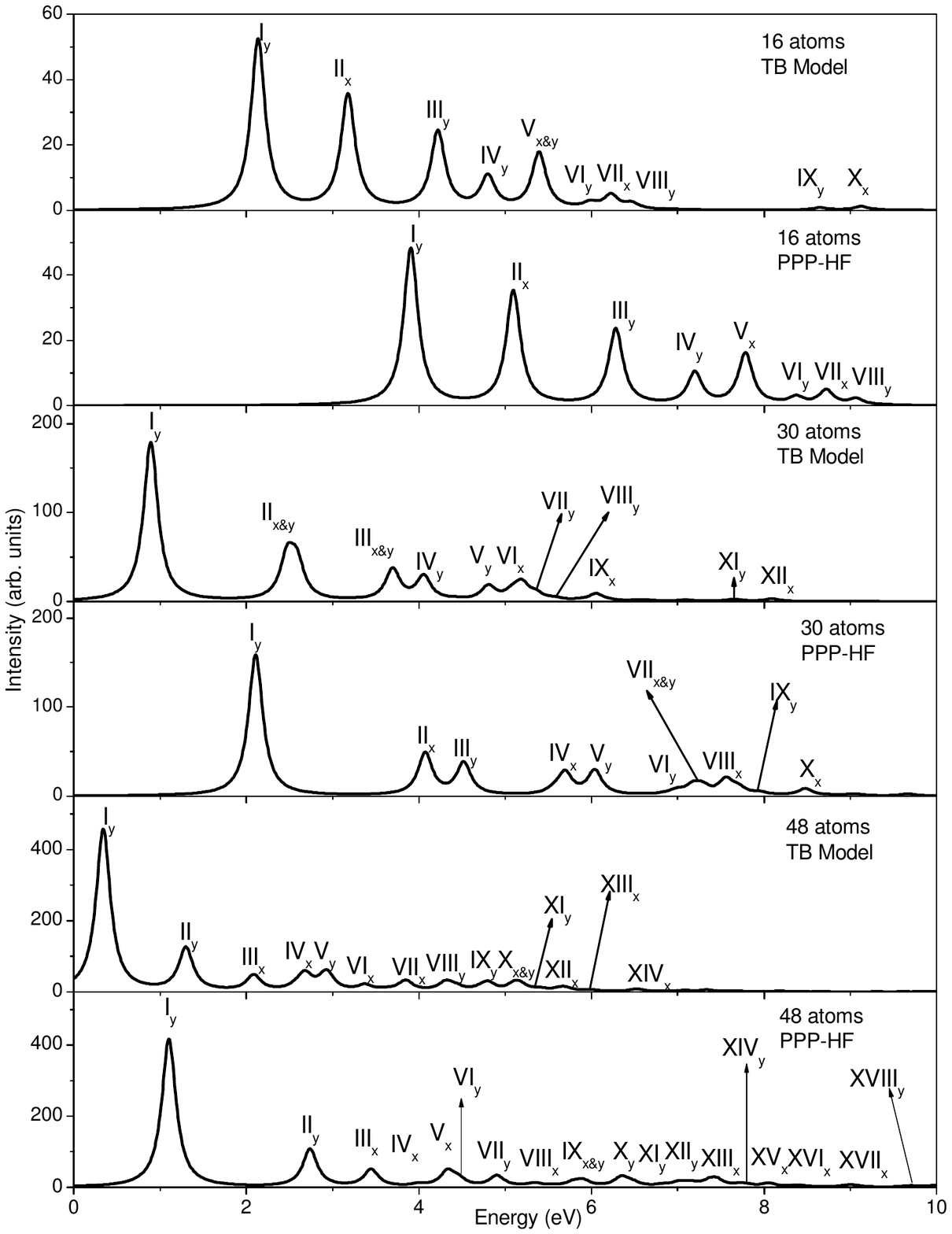}\protect

\vspace{-4cm}

\caption{Optical absorption spectrum of DQD-16, DQD-30, and DQD-48, calculated
using the tight binding model and PPP model at HF level employing
screened parameters. The spectrum has been broadened with a uniform
line width of 0.1 eV. \label{fig:Computed-optical-absorption}}
\end{figure}

\par\end{flushleft}

\begin{table}
\caption{Comparison of locations of experimentally measured linear absorption
peaks of pyrene and dibenzo{[}bc,kl{]}coronene with our PPP-CI results
for DQD-16 and DQD-30, respectively. Computed values of the excited
state energies for DQD-48 are also presented. DF and MI, respectively,
denote dipole forbidden state, and most intense peak. Theoretical
results of other authors are also presented for comparison. All energies
are in eV units.}

\centering{}%
\begin{tabular}{|c|c|c|c|cc|}
\hline 
{\footnotesize{}System} & {\footnotesize{}Symmetry} & {\footnotesize{}Experimental Values} & {\footnotesize{}Theory } & \multicolumn{2}{c|}{{\footnotesize{}This work}}\tabularnewline
 &  &  & {\footnotesize{}(Other)} & \multicolumn{2}{c|}{}\tabularnewline
\hline 
 &  &  &  & {\footnotesize{}Scr} & {\footnotesize{}Std}\tabularnewline
\hline 
{\footnotesize{}DQD-16} & {\footnotesize{}$B$$_{3u}$ (DF)} & {\footnotesize{}3.34,}\cite{pyrene-exp-indrasen} {\footnotesize{}3.33,}\cite{Gudipati_1993}  & {\footnotesize{}3.33,\cite{Canuto} 3.83,\cite{Gudipati_1993} 3.75}\cite{Parac200311}  & {\footnotesize{}2.82} & {\footnotesize{}3.09 }\tabularnewline
\hline 
 & {\footnotesize{}$B$$_{2u}$} & {\footnotesize{}3.70,}\cite{pyrene-exp-indrasen} {\footnotesize{}3.69},\cite{BasuRay2006248,pyrene-photocad}
{\footnotesize{}3.71},\cite{Ram20092252} {\footnotesize{}3.80},\cite{Salama_1993}  & {\footnotesize{}3.35},\cite{pyrene-theory-malloci} {\footnotesize{}3.53},\cite{Canuto}
{\footnotesize{}3.93},\cite{Gudipati_1993} \textcolor{red}{\footnotesize{}3.69}\cite{Parac200311} & {\footnotesize{}3.60 } & {\footnotesize{}3.74}\tabularnewline
 & {\footnotesize{}(optical gap)} & {\footnotesize{}3.75},\cite{pyrene-exp-vala} {\footnotesize{}3.79},\cite{Gudipati_1993}
{\footnotesize{}3.83}\cite{Halasinski_2005} &  &  & \tabularnewline
\hline 
 & {\footnotesize{}$B$$_{3u}$} & {\footnotesize{}4.55},\cite{BasuRay2006248,Ram20092252,pyrene-exp-indrasen}
{\footnotesize{}4.62},\cite{Gudipati_1993}   & {\footnotesize{}4.10},\cite{pyrene-theory-malloci} {\footnotesize{}4.70},\cite{Canuto}
{\footnotesize{}5.26}\cite{Gudipati_1993} & {\footnotesize{}4.37} & {\footnotesize{}4.94}\tabularnewline
 &  & {\footnotesize{}4.67},\cite{Salama_1993,Halasinski_2005} {\footnotesize{}4.60},\cite{pyrene-exp-vala} &  &  & \tabularnewline
\hline 
 & {\footnotesize{}$B$$_{2u}$} & {\footnotesize{}5.15,}\cite{pyrene-exp-indrasen} {\footnotesize{}5.12},\cite{BasuRay2006248,pyrene-photocad}
{\footnotesize{}5.17},\cite{Ram20092252} {\footnotesize{}5.29},\cite{Gudipati_1993}  & {\footnotesize{}5.00},\cite{pyrene-theory-malloci} {\footnotesize{}5.36}
,\cite{Canuto} {\footnotesize{}5.60} \cite{Gudipati_1993} & {\footnotesize{}5.37} & {\footnotesize{}5.44 }\tabularnewline
 &  & {\footnotesize{}5.34},\cite{Salama_1993,Halasinski_2005} {\footnotesize{}5.35},\cite{Thony}
{\footnotesize{}5.22}\cite{pyrene-exp-vala} &  &  & \tabularnewline
\hline 
 & {\footnotesize{}$B$$_{2u}$} & {\footnotesize{}6.32},\cite{pyrene-exp-indrasen} {\footnotesize{}6.07}\cite{Gudipati_1993} & {\footnotesize{}5.80},\cite{pyrene-theory-malloci} {\footnotesize{}5.96},\cite{Canuto}
{\footnotesize{}6.66}\cite{Gudipati_1993} & {\footnotesize{}6.22} & {\footnotesize{}6.44 }\tabularnewline
\hline 
 & {\footnotesize{}$B$$_{3u}$} & {\footnotesize{}6.42}\cite{Gudipati_1993} & {\footnotesize{}6.04},\cite{Canuto} {\footnotesize{}6.35} ,\cite{pyrene-theory-malloci}
{\footnotesize{}6.96}\cite{Gudipati_1993} & {\footnotesize{}6.38} & {\footnotesize{}6.96}\tabularnewline
\hline 
 & {\footnotesize{}$B$$_{2u}$, $B$$_{3u}$} & {\footnotesize{}7.02}\cite{Gudipati_1993} & {\footnotesize{}6.45},\cite{Canuto} {\footnotesize{}7.39},\cite{pyrene-theory-malloci}
{\footnotesize{}7.51}\cite{Gudipati_1993} & {\footnotesize{}6.88, 6.87} & {\footnotesize{}7.29, 7.33}\tabularnewline
\hline 
 & {\footnotesize{}$B$$_{2u}$(MI)} & {\footnotesize{}5.15},\cite{pyrene-exp-indrasen} {\footnotesize{}5.12},\cite{BasuRay2006248,pyrene-photocad}
{\footnotesize{}5.17},\cite{Ram20092252} {\footnotesize{}5.29},\cite{Gudipati_1993}  & {\footnotesize{}6.35},\cite{pyrene-theory-malloci} {\footnotesize{}5.36}
,\cite{Canuto} {\footnotesize{}5.60} \cite{Gudipati_1993} & {\footnotesize{}5.37} & {\footnotesize{}6.44}\tabularnewline
 &  & {\footnotesize{}5.34},\cite{Salama_1993,Halasinski_2005} {\footnotesize{}5.35},\cite{Thony}
{\footnotesize{}5.22}\cite{pyrene-exp-vala} &  &  & \tabularnewline
\hline 
{\footnotesize{}DQD-30} & {\footnotesize{}$B$$_{2u}$} & {\footnotesize{}2.55}\cite{clar-dbcoronene}  & {\footnotesize{}2.10}\cite{dibenzocoronene-database,pah-database-malloci} & {\footnotesize{}2.08} & {\footnotesize{}2.31}\tabularnewline
 & {\footnotesize{}(optical gap)} &  &  &  & \tabularnewline
\hline 
 & {\footnotesize{}$B$$_{3u}$ (DF)} &  &  & {\footnotesize{}2.25} & {\footnotesize{}2.43}\tabularnewline
\hline 
 & {\footnotesize{}$B$$_{3u}$} &  {\footnotesize{}3.61}\cite{clar-dbcoronene} & {\footnotesize{}3.55}\cite{dibenzocoronene-database,pah-database-malloci} & {\footnotesize{}3.45} & {\footnotesize{}3.83}\tabularnewline
\hline 
 & {\footnotesize{}$B$$_{3u}$(MI)} & {\footnotesize{}6.20}\cite{dbcoronene-database}  & {\footnotesize{}5.90}\cite{dibenzocoronene-database,pah-database-malloci} & 6.28  & 5.10 \tabularnewline
\hline 
{\footnotesize{}DQD-48} & {\footnotesize{}$B$$_{3u}$ (DF)} &  &  & 1.38 & 1.52\tabularnewline
\hline 
 & {\footnotesize{}$B$$_{2u}$} &  &  & 1.40 & 1.61\tabularnewline
 & {\footnotesize{}(optical gap)} &  &  &  & \tabularnewline
\hline 
 & {\footnotesize{}$B$$_{2u}$(MI)} &  &  & {\footnotesize{}2.19} & {\footnotesize{}5.54}\tabularnewline
\hline 
\end{tabular}\label{tab:exp-pyrene}
\end{table}

\subsubsection{Correlated-electron calculations}

We discuss the general features of the optical absorption spectra
obtained from the CI calculations, followed by a more detailed examination
of individual DQDs. 

One observes the following general trends upon examining the absorption
spectra calculated by the PPP-CI approach presented in Figs. \ref{fig:Computed-linear-optical-16-atoms}---\ref{fig:Computed-linear-absorption-48-atoms},
and the quantitative information about various excited states detailed
in Tables \ref{tab:Linear 16 atoms standard-1}---\ref{tab:Linear absorption 48 atoms screened}
of the Appendix: 
\begin{enumerate}
\item In agreement with the TB results, the absorption spectrum is red-shifted
with the increasing size of the DQD. 
\item The spectra obtained from screened parameters is red-shifted as compared
to that obtained from standard parameters. Furthermore, for the case
of DQD-16\cite{pyrene-exp-indrasen,pyrene-photocad,pyrene-exp-vala,BasuRay2006248,Gudipati_1993,Halasinski_2005,Ram20092252,Salama_1993,Thony}
and\textcolor{red}{{} }DQD-30\cite{dbcoronene-database,clar-dbcoronene},
screened parameter results are in overall better agreement with the
experimental data (\emph{cf}. Table \ref{tab:exp-pyrene}), as compared
to the standard parameter ones.
\item The first peak in the spectrum of the DQDs is always due to the absorption
of a $y$-polarized photon, causing a transition from their ground
state ($1^{1}A_{g}$), to the $1^{1}B_{2u}$ excited state, and denotes
the optical gap. The wave function of the $1^{1}B_{2u}$ state for
all the DQDs is dominated by the $|H\rightarrow L\rangle$ excitation,
in agreement with the results of the TB model. However, a quantitative
analysis of the optical gap indicates that its value obtained from
the TB model is much less compared to the value obtained from the
PPP-CI approach. For the cases of DQD-16\cite{pyrene-exp-indrasen,pyrene-photocad,pyrene-exp-vala,BasuRay2006248,Gudipati_1993,Halasinski_2005,Ram20092252,Salama_1993,Thony}
and DQD-30,\cite{dbcoronene-database,clar-dbcoronene} for which the
experimental results are available, again PPP-CI value of the gap
is in much better agreement with experimental results, than the TB
model value. Therefore, we hope that similar experiments can be performed
on DQD-48 in the future, so that our predicted PPP-CI values of optical
gaps can be tested. 
\item The intensity of the first peak is lesser as compared to the intensity
of other higher energy peak(s) in the spectrum, in contrast to the
predictions of the TB model, and in agreement with the experimental
results for pyrene,\cite{pyrene-exp-indrasen,pyrene-photocad,pyrene-exp-vala,BasuRay2006248,Gudipati_1993,Halasinski_2005,Ram20092252,Salama_1993,Thony}
and dibenzo{[}bc,kl{]}coronene.\cite{dbcoronene-database,clar-dbcoronene}
\item The calculated position of the first peak in the PPP-CI absorption
spectrum is weakly dependent on the choice of the Coulomb parameters
in the PPP model (standard or screened). However, higher energy peaks,
and the character of the many-particle wave functions contributing
to them, do depend significantly upon the choice of Coulomb parameters.
\textcolor{black}{In particular, the position of the most intense
peak is drastically dependent upon the choice of the Coulomb parameters
in the PPP-CI calculations.}\textcolor{green}{{} }Thus, we conclude
that the position of the most intense peak is strongly dependent on
the strength of the Coulomb interactions in theses systems.
\item The optical transition to the first excited state of $B_{3u}$ symmetry
is dipole forbidden within the PPP model, on account of the particle-hole
symmetry due to the use of the nearest-neighbour hopping approximation.
However, this symmetry is approximate in the real systems, and hence
the transition to this state is experimentally observed as a weak
peak. Our PPP-CI calculations predict this state to lie below the
optical gap for DQD-16 and DQD-48, but above it for DQD-30. (\emph{cf}.
Tables \ref{tab:Linear 16 atoms standard-1}, \ref{tab:Linear absorption 16 atoms screened-1},
\ref{tab:Linear absorption 30 atoms standard}, \ref{tab:Linear absorption 30 atoms screened},
\ref{tab:Linear absorption 48 atoms standard-1} and \ref{tab:Linear absorption 48 atoms screened}).
Our prediction is in agreement with the experiments for the case of
DQD-16 when compared with the data for pyrene (\emph{cf}. Table \ref{tab:exp-pyrene}
), however, no experimental results for this state are available for
the larger DQDs.
\item Wave functions of higher energy states derive significant contributions
from double and higher level excitations, signaling the importance
of electron correlation effects.
\end{enumerate}

\subsubsection*{DQD-16}

\begin{figure}
\includegraphics[width=8cm]{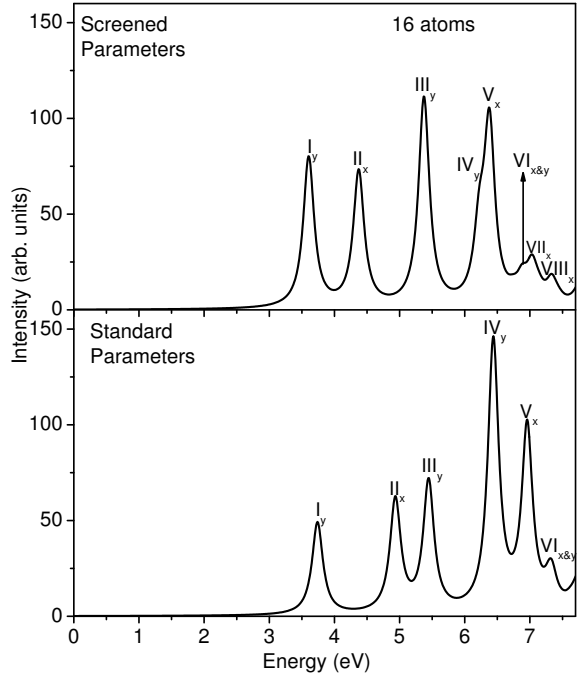}\protect\caption{Computed linear optical absorption spectrum for \label{fig:Computed-linear-optical-16-atoms}16
atoms DQD, obtained by employing screened as well as standard parameters.
In both the cases, the spectrum has been broadened with a uniform
line-width of 0.1 eV.}
\end{figure}

Figure \ref{fig:Computed-linear-optical-16-atoms} presents the computed
linear absorption spectrum for DQD-16, obtained by employing standard
as well as screened parameters, while Tables \ref{tab:Linear 16 atoms standard-1}
and \ref{tab:Linear absorption 16 atoms screened-1}, of the Appendix,
present the detailed quantitative data corresponding to various peaks
in the computed spectra, and the excited states contributing to them.
Our theoretical results have been compared with the experimental optical
absorption data of pyrene (C$_{16}$H$_{10}$), which is nothing but
hydrogen saturated DQD-16. Because we have employed PPP model parameters
used to describe the optical properties of aromatic hydrocarbons,
the comparison between DQD-16 and pyrene is most appropriate. A number
of theoretical\cite{Canuto,Gudipati_1993,Parac200311,pyrene-theory-malloci}
and experimental studies\cite{pyrene-exp-indrasen,pyrene-photocad,pyrene-exp-vala,BasuRay2006248,Gudipati_1993,Halasinski_2005,Ram20092252,Salama_1993,Thony}
of optical absorption in pyrene have been carried out in the past,
and our calculated excited state energies are found to to be in very
good agreement with the results obtained earlier (\emph{cf}. Table
\ref{tab:exp-pyrene} ). 

The first peak in the experimentally obtained absorption spectrum
of pyrene is a weak one located around 3.34 eV,\cite{pyrene-exp-indrasen}
and corresponds to the dipole forbidden $B_{3u}$ state in our calculations,
mentioned earlier. Our standard parameter value of 3.09 eV for the
excitation energy of this state is in good agreement with the experimental
value, while the screened parameter value of 2.82 eV underestimates
it.\textcolor{red}{{} }The location of the second peak in the experimental
spectrum (3.69-3.83 eV), which also defines the optical gap, is in
excellent agreement both with our standard and screened parameter
PPP-CI values of the optical gap, computed at 3.74 eV, and 3.60 eV,
respectively. This peak is $y$-polarized and corresponds to $1B_{2u}$
state. The optical transition to the fourth excited state gives rise
to the most intense peak experimentally observed to be in the range
5.15--5.35 eV. This result is in excellent agreement with our PPP-CI
value 5.37 eV, obtained using the screened parameters.\textcolor{red}{{}
}As a matter of fact, it is obvious from Table \ref{tab:exp-pyrene},
that the agreement between the dipole-allowed states obtained from
our screened-parameter based PPP-CI calculations, and the experimental
measurements of Becker \emph{et al}.\cite{pyrene-exp-indrasen} and
Gudipati \emph{et al}.\cite{Gudipati_1993} is quite remarkable both
for peak locations, and the symmetry assignments, all the way up to
7 eV. On the other hand, the PPP-CI results obtained using standard
parameters, as also the earlier results obtained by Malloci \emph{et
al}.,\cite{pyrene-theory-malloci} predict that transition to the
fifth excited state gives rise to the most intense peak. Thus, we
conclude that, on the whole, the PPP-CI results calculated using the
screened parameters are in better agreement with the experimental
values than those computed using the standard parameters (\emph{cf}.
Table \ref{tab:exp-pyrene}).  Screened parameter based calculations
also predict that the wave functions of the excited states of the
first five peaks (I--V) are dominated by single excitations (\emph{cf}.
Table \ref{tab:Linear absorption 16 atoms screened-1}).  We also
note that the TB model predicts the peak corresponding to the optical
gap as the most intense one, located at 2.14 eV, which is far away
from the experimentally obtained value both in terms of peak location,
and relative intensity. Therefore, we infer that the inclusion of
electron correlation effects is essential for the correct quantitative
description of the optical properties of graphene quantum dots. 

\subsubsection*{DQD-30 and DQD-48}

\begin{figure}
\includegraphics[width=8cm]{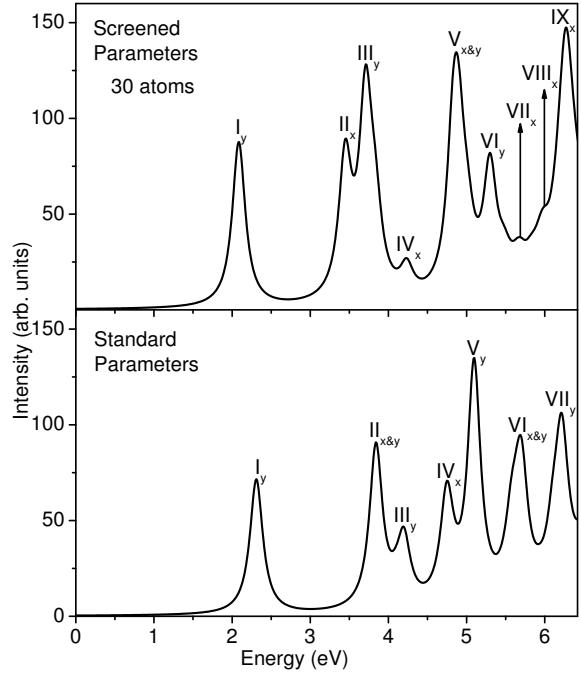}\protect\caption{\label{fig:Computed-linear-optical-30-atoms}Computed linear optical
absorption spectrum for 30 atoms DQD, obtained by employing screened
as well as standard parameters. In both the cases, the spectrum has
been broadened with a uniform line-width of 0.1 eV.}
\end{figure}

\begin{figure}
\includegraphics[width=8cm]{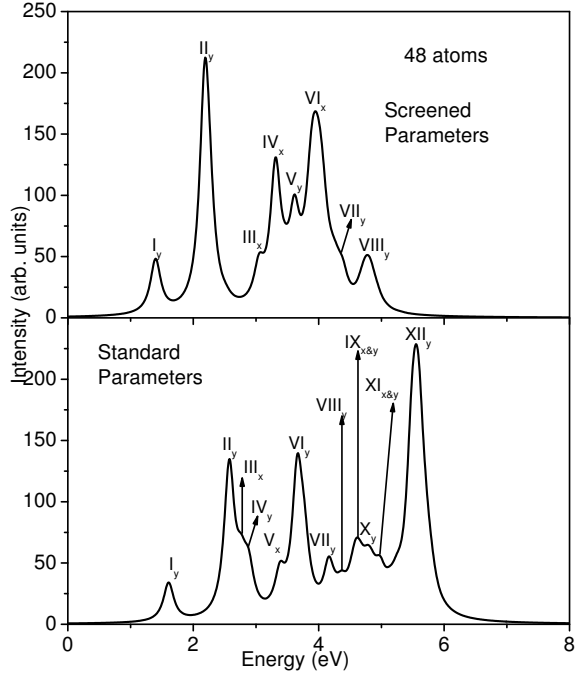}\protect\caption{\label{fig:Computed-linear-absorption-48-atoms}Computed linear absorption
spectra (with a line-width of 0.1 eV) for 48 atoms DQD, obtained by
employing standard as well as screened parameters}
\end{figure}

In Figs. \ref{fig:Computed-linear-optical-30-atoms} and \ref{fig:Computed-linear-absorption-48-atoms},
we present the computed linear absorption spectra for DQD-30 and DQD-48,
respectively. Information related to the energies, transition dipoles,
and many-particle wave functions of excited states contributing to
various absorption peaks for DQD-30 are presented in Tables \ref{tab:Linear absorption 30 atoms standard}
and \ref{tab:Linear absorption 30 atoms screened} of the Appendix.

As far as DQD-30 is concerned, our computed absorption spectrum has
been compared with the experimental data of dibenzo{[}bc,kl{]}coronene
(C$_{30}$H$_{14}$).\cite{clar-dbcoronene} The experimental UV spectrum
obtained by Clar and Schmidt,\cite{clar-dbcoronene} exhibits peaks
at 2.55 eV and 3.61 eV. The position of the first peak at 2.55 eV
is in good agreement with the computed value of optical gap at 2.08
eV (2.31 eV) (\emph{cf}. Table \ref{tab:exp-pyrene}) obtained using
screened (standard) parameters in the PPP-CI model. This peak is $y$-polarized,
and corresponds to the $1B_{2u}$ state. The first excited state of
$B_{3u}$ symmetry is dipole forbidden due to the particle-hole symmetry,
and lies above the optical gap. Its energy is 2.25 eV (2.43 eV) obtained
using screened (standard) parameters and is dominated by the $|H-2\rightarrow L\rangle+c.c.$
excitation. In addition, the experimental peak at 3.61 eV agrees extremely
well with the screened parameter $x-$polarized peak at 3.45 eV. This
peak corresponds to a $B_{3u}$ state, whose wave function is dominated
by the $|H-2\rightarrow L\rangle-c.c$ excitation. Thus, same excitations
contribute to the wave functions of the first dipole-forbidden and
dipole-allowed $B_{3u}$ states, it is just that their relative signs
are opposite due to the orthogonality constraint. Further, the most
intense peak of the experimental spectrum is situated at around 6.20
eV,\cite{dbcoronene-database} which is in excellent agreement with
the computed screened parameter value of the most intense peak at
6.28 eV. This peak also corresponds to a $B_{3u}$ state whose wave
function consists of singly as well as doubly excited configurations
(\emph{cf}. Table \ref{tab:Linear absorption 30 atoms screened}).
On the other hand, our standard parameter calculations predict the
most intense absorption peak at 5.10 eV corresponding to a $B_{2u}$
state, whose wave function consists mainly of single excitations,
dominated by the configuration $|H-2\rightarrow L+2\rangle$ (\emph{cf}.
Table \ref{tab:Linear absorption 30 atoms standard}). Thus, our computations
imply that screened parameter values are in better overall agreement
with the experimental results than standard parameter values, and
TB model predictions. While wave functions of the excited states corresponding
to various peaks are dominated by single excitations, however, several
states also derive significant contributions from the double excitations,
hinting at the importance of electron correlation effects. 

In case of DQD-48, because of comparatively larger number of electrons
in the system, the size of the MRSDCI calculations became excessively
large. Therefore, we froze two lowest lying occupied orbitals, and
deleted their particle-hole counterpart virtual orbitals which were
highest in energy. With this approximation in place, the CI problem
reduced to that of 44 electrons, distributed over 22 occupied, and
as many virtual orbitals, rendering the calculation tractable. In
order to benchmark this procedure, we also adopted the same methodology
for DQD-30, and present the results of the calculations performed
using screened parameters in Fig. \ref{fig:Computed-absorption-spectrum-freeze}
of the Appendix. It is observed that all the features of the optical
spectra are preserved even after freezing the orbitals. However, the
frozen spectrum is slightly blue-shifted as compared to the unfrozen
one, with the corresponding changes being numerically acceptable.
Next, we discuss our results for DQD-48 presented in Fig. \ref{fig:Computed-linear-absorption-48-atoms},
and Tables \ref{tab:Linear absorption 48 atoms standard-1}, \ref{tab:Linear absorption 48 atoms standard},
and \ref{tab:Linear absorption 48 atoms screened}.

We find that the first excited state of DQD-48 is a dipole forbidden
$B_{3u}$ state, just as in the case of DQD-16, and is located at
1.38 eV (1.52 eV) as per our screened (standard) parameter calculations.
Both the calculations predict it to be lower than the first dipole
allowed state $1B_{2u}$, although the energy difference is much smaller
as compared to the case of DQD-16. The wave function of this $B_{3u}$
state is dominated by the double excitation $|H\rightarrow L+1;H\rightarrow L\rangle+c.c.$,
and it will be of considerable interest if the future experiments
on DQD-48 are able to locate this state relative to the optical gap.
The first dipole-allowed peak in the absorption spectrum of DQD-48,
corresponding to the $1B_{2u}$ state of the spectrum as in case of
smaller DQDs, is computed at 1.61 eV (1.40 eV) based upon standard
(screened) parameter based PPP-CI calculations. The wave function
of this state is dominated by the $|H\rightarrow L\rangle$ excitation
as in case of smaller dots, but it also derives significant contribution
from the $|H-1\rightarrow L+1\rangle$ configuration. The most intense
peak of the absorption spectrum computed with the screened parameter
is peak II corresponding to the $2B_{2u}$ state located at 2.19 eV,
with the wave function dominated by the $|H-1\rightarrow L+1\rangle$
configuration, but also with a significant contribution from a triply
excited configuration. On the other hand, standard parameter calculations
predict peak XII to be the most intense one, which is due to a high-energy
$B_{2u}$ state at 5.54 eV, with the wave function dominated by several
configurations. Such a large difference in the locations of the most
intense peak predicted by standard and screened parameter calculations,
can be easily tested in experiments. As far as general comparison
between the standard and screened parameter calculations is concerned,
besides the red shift of the screened parameter results compared to
the standard ones, we find that only the first two peaks of the computed
spectra have excited state wave functions which are qualitatively
similar. We also note that compared to the two smaller DQDs discussed
earlier, DQD-48 excited states exhibit significantly more contribution
from doubly excited configurations. This trend implies higher contribution
of electron-correlation effects in DQD-48. 

\section{Conclusions}

\label{sec:conclusions}

In this work, very large scale correlated calculations employing the
PPP Hamiltonian were carried out on diamond shaped graphene quantum
dots of increasing sizes, namely, DQD-16, DQD-30, and DQD-48, and
their optical as well as electronic properties were computed. Calculated
linear optical absorption spectra of DQD-16 and DQD-30 were found
to be in very good agreement with the experimental data of pyrene
(C\textsubscript{\textcolor{black}{16}}H$_{10}$) and dibenzo{[}bc,kl{]}coronene
(C\textsubscript{30}H$_{14}$), which are their respective structural
analogs with hydrogen passivated edges, thus justifying the essential
correctness of our methodology. Some of the important conclusions
we can draw from our correlated-electron calculations are: (i) the
first peak corresponding to the optical gap is not the most intense,
in contrast with the predictions of the tight-binding model, (ii)
with the increasing size of the quantum dot, the absorption spectrum
exhibits a red shift, (iii) the optical transition to the first excited
state of $B_{3u}$ symmetry is dipole forbidden and it lies below
the optical gap for DQD-16 and DQD-48, (iv) optical properties of
the dots are sensitive to the projected corners of the system, therefore,
they can be tuned by attaching suitable functional groups there. Thus,
we hope that our work will spur further experimental activity in this
field, so that our predictions on the excited states of DQD-48 can
be tested in future experiments. Furthermore, recently M\"ullen and
coworkers have stabilized graphene quantum dots with chlorine passivated
edges.\cite{Tan_GNR} Therefore, it will be of interest if chlorine
passivated DQDs can be synthesized and their optical properties measured,
so as to investigate the influence of the nature of edge passivation
on the electro-optical properties of graphene nanostructures.

In this work, we restricted ourselves to the study of linear optical
properties of these quantum dots, but it will be quite interesting
also to study the nonlinear optical response of these systems such
as two-photon absorption, third harmonic generation, and photoinduced
absorption. Calculations along those directions are underway in our
group, and the results will be submitted for publication in future.
\begin{acknowledgments}
One of us (H.C.) thanks Council of Scientific and Industrial Research
(CSIR), India for providing a Senior Research Fellowship (SRF). We
thankfully acknowledge the computational resources (PARAM-YUVA) provided
for this work by Center for Development of Advanced Computing (C-DAC),
Pune. We also gratefully acknowledge the financial support of  Department
of Science and Technology (DST), India, under the grant \# SB/S2/CMP-066/2013.
\end{acknowledgments}

\appendix

\section{Influence of orbital freezing and deletion on the optical absorption
spectrum of DQD-30}

In order to benchmark our orbital freezing and deletion approach aimed
at reducing the size of the CI calculations, in the figure below we
present the results of optical absorption spectra of DQD-30 computed
using all the electrons and orbitals, and by freezing (deleting) two
lowest (highest) occupied (virtual) orbitals. Apart from a slight
blue shift in the frozen orbital calculation, computed spectra are
quite similar.

\begin{figure}
\includegraphics[scale=0.5]{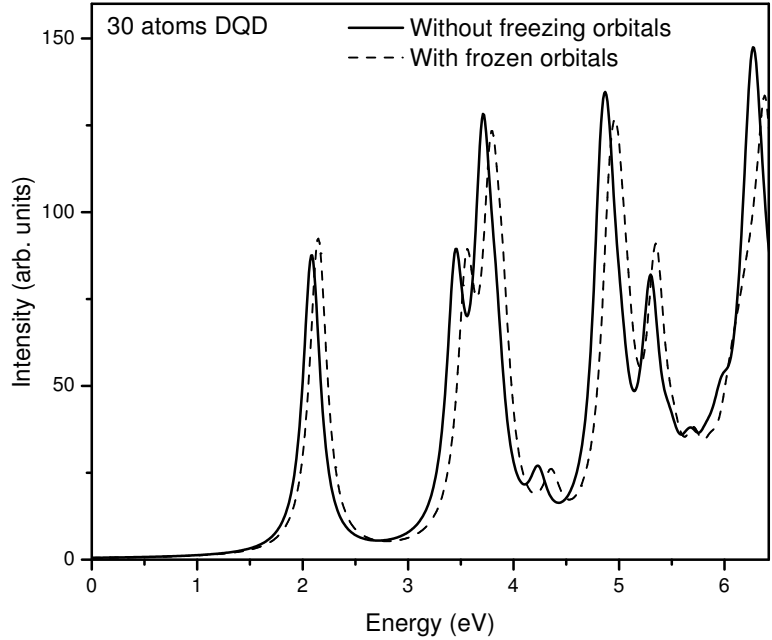}\protect\caption{Computed absorption spectrum \label{fig:Computed-absorption-spectrum-freeze}
(with a line-width of 0.1 eV) for 30 atoms DQD, obtained by freezing
orbitals (dotted line) and without freezing orbitals (bold line).}
\end{figure}

\section{Calculated Energies, Wave Functions, and Transition Dipole Moments
Of The Excited States Contributing To The Linear Absorption Spectra}

Following tables represent the excitation energies, dominant many-body
wave-functions, and transition dipole matrix elements of excited states
with respect to the ground state (1$^{1}$A$_{g}$). The coefficient
of charge conjugate of a given singly excited configuration is abbreviated
as 'c.c.', while the sign (+/-) preceding 'c.c.' indicates that the
two coefficients have (same/opposite) signs. For the doubly excited,
and the higher order configurations, no +/- sign precedes c.c. because
more than one charge-conjugate counterparts are possible, each with
its own sign. Label DF associated with a peak implies that the excited
state in question is dipole forbidden.

\begin{table}
\protect\caption{Excited states giving rise to the peaks in the linear absorption spectrum
of 16 atoms DQD, computed employing the QCI ($B_{2u}$ states) and
MRSDCI ($B_{3u}$ states) approaches, along with the standard parameters
in the PPP model Hamiltonian. }

\begin{tabular}{|c|c|c|c|c|}
\hline 
{\small{}Peak } & {\small{}State } & {\small{}E (eV) } & {\small{}Transition Dipole (Å) } & {\small{}Dominant Configurations}\tabularnewline
 &  &  &  & \tabularnewline
\hline 
{\small{}$DF$} & {\small{}$1^{1}B\textrm{\ensuremath{_{3u}}}$ } & {\small{}$3.09$} & {\small{}$0$} & {\small{}$|H\rightarrow L+1\rangle$$-c.c.(0.5919)$ }\tabularnewline
 &  &  &  & {\small{}$|H-4\rightarrow L+3\rangle$$-c.c.(0.1244)$ }\tabularnewline
\hline 
{\small{}$I_{y}$ } & {\small{}$1^{1}B\textrm{\ensuremath{_{2u}}}$ } & {\small{}$3.74$ } & {\small{}$1.137$ } & {\small{}$|H\rightarrow L\rangle$$(0.8606)$}\tabularnewline
 &  &  &  & {\small{}$|H-1\rightarrow L+1\rangle$$(0.2855)$}\tabularnewline
\hline 
{\small{}$II_{x}$ } & {\small{}$2^{1}B\textrm{\ensuremath{_{3u}}}$ } & {\small{}$4.94$ } & {\small{}$1.094$ } & {\small{}$|H-1\rightarrow L\rangle$$+c.c.(0.5897)$}\tabularnewline
 &  &  &  & {\small{}$|H\rightarrow L+1;H-3\rightarrow L\rangle$$c.c.(0.1440)$}\tabularnewline
 &  &  &  & {\small{}$|H-2\rightarrow L+3\rangle$$-c.c.(0.1377)$}\tabularnewline
\hline 
{\small{}$III_{y}$ } & {\small{}$3^{1}B\textrm{\ensuremath{_{2u}}}$ } & {\small{}$5.44$ } & {\small{}$1.117$ } & {\small{}$|H-1\rightarrow L+1\rangle$$(0.7593)$}\tabularnewline
 &  &  &  & {\small{}$|H-2\rightarrow L+2\rangle$$(0.3370)$}\tabularnewline
\hline 
{\small{}$IV_{y}$ } & {\small{}$5^{1}B\textrm{\ensuremath{_{2u}}}$ } & {\small{}$6.44$ } & {\small{}$1.482$ } & {\small{}$|H-2\rightarrow L+2\rangle$$(0.7125)$}\tabularnewline
 &  &  &  & {\small{}$|H-1\rightarrow L+1\rangle$$(0.3400)$}\tabularnewline
\hline 
{\small{}$V_{x}$ } & {\small{}$8^{1}B\textrm{\ensuremath{_{3u}}}$ } & {\small{}$6.96$ } & {\small{}$1.169$ } & {\small{}$|H-3\rightarrow L+2\rangle$$-c.c.(0.4736)$}\tabularnewline
 &  &  &  & {\small{}$|H\rightarrow L+2;H\rightarrow L\rangle$$+c.c.(0.3047)$}\tabularnewline
\hline 
{\small{}$VI_{x\&y}$ } & {\small{}$10B\textrm{\ensuremath{_{2u}}}$ } & {\small{}$7.29$ } & {\small{}$0.255$ } & {\small{}$|H-2\rightarrow L+4\rangle$$-c.c.(0.4346)$}\tabularnewline
 &  &  &  & {\small{}$|H-3\rightarrow L+3\rangle$$(0.3309)$}\tabularnewline
\hline 
 & {\small{}$9^{1}B\textrm{\ensuremath{_{3u}}}$ } & {\small{}$7.33$ } & {\small{}$0.459$ } & {\small{}$|H\rightarrow L+6\rangle$$-c.c.(0.4957)$}\tabularnewline
 &  &  &  & {\small{}$|H-1\rightarrow L;H-3\rightarrow L\rangle$$-c.c.(0.1993)$}\tabularnewline
\hline 
\end{tabular}\label{tab:Linear 16 atoms standard-1}
\end{table}

\begin{table}
\protect\caption{Excited states giving rise to the peaks in the linear absorption spectrum
of 16 atoms DQD, computed employing the QCI ($B_{2u}$ states) and
MRSDCI ($B_{3u}$ states) approaches, along with the screened parameters
in the PPP model Hamiltonian. }

\begin{tabular}{|c|c|c|c|c|}
\hline 
{\small{}Peak } & {\small{}State } & {\small{}E (eV) } & {\small{}Transition Dipole (Å) } & {\small{}Dominant Configurations}\tabularnewline
 &  &  &  & \tabularnewline
\hline 
{\small{}$DF$} & {\small{}$1^{1}B\textrm{\ensuremath{_{3u}}}$ } & {\small{}$2.82$} & {\small{}$0$.000} & {\small{}$|H\rightarrow L+1\rangle$$-c.c.(0.5730)$ }\tabularnewline
 &  &  &  & {\small{}$|H\rightarrow L+4;H\rightarrow L\rangle$$+c.c.(0.1090)$}\tabularnewline
\hline 
{\small{}$I_{y}$ } & {\small{}$1^{1}B\textrm{\ensuremath{_{2u}}}$ } & {\small{}$3.60$ } & {\small{}$1.477$ } & {\small{}$|H\rightarrow L\rangle$$(0.8802)$}\tabularnewline
\hline 
{\small{}$II_{x}$ } & {\small{}$2^{1}B\textrm{\ensuremath{_{3u}}}$ } & {\small{}$4.37$ } & {\small{}$1.272$ } & {\small{}$|H-1\rightarrow L\rangle$$+c.c.$ $(0.6042)$ }\tabularnewline
 &  &  &  & {\small{}$|H\rightarrow L+1;H-3\rightarrow L\rangle$$c.c.$ $(0.0820)$}\tabularnewline
 &  &  &  & {\small{}$|H-3\rightarrow L+2\rangle$$+c.c.$ $(0.0818)$}\tabularnewline
\hline 
{\small{}$III_{y}$ } & {\small{}$3^{1}B\textrm{\ensuremath{_{2u}}}$ } & {\small{}$5.37$ } & {\small{}$1.424$ } & {\small{}$|H-1\rightarrow L+1\rangle$$(0.8596)$}\tabularnewline
 &  &  &  & {\small{}$|H-1\rightarrow L+1;H\rightarrow L;H\rightarrow L\rangle$$(0.1113)$}\tabularnewline
\hline 
{\small{}$IV_{y}$ } & {\small{}$5^{1}B\textrm{\ensuremath{_{2u}}}$ } & {\small{}$6.22$ } & {\small{}$0.746$ } & {\small{}$|H-2\rightarrow L+2\rangle$$(0.6716)$ }\tabularnewline
 &  &  &  & {\small{}$|H\rightarrow L+5\rangle$$-c.c.$ $(0.3187)$}\tabularnewline
\hline 
{\small{}$V_{x}$ } & {\small{}$8^{1}B\textrm{\ensuremath{_{3u}}}$ } & {\small{}$6.38$ } & {\small{}$1.205$ } & {\small{}$|H-2\rightarrow L+3\rangle$$+c.c.(0.5243)$}\tabularnewline
 &  &  &  & {\small{}$|H\rightarrow L+2;H\rightarrow L\rangle$$+c.c.(0.2700)$}\tabularnewline
\hline 
{\small{}$VI_{x\&y}$ } & {\small{}$9^{1}B\textrm{\ensuremath{_{3u}}}$ } & {\small{}$6.87$ } & {\small{}$0.049$ } & {\small{}$|H\rightarrow L+6\rangle$$-$$c.c.(0.3507)$}\tabularnewline
 &  &  &  & {\small{}$|H-4\rightarrow L+3\rangle$$-c.c.(0.2767)$}\tabularnewline
 & {\small{}$10^{1}B\textrm{\ensuremath{_{2u}}}$ } & {\small{}$6.88$ } & {\small{}$0.396$ } & {\small{}$|H-3\rightarrow L+3\rangle$$(0.3983)$}\tabularnewline
 &  &  &  & {\small{}$|H-2\rightarrow L+4\rangle$$-c.c.(0.2996)$}\tabularnewline
 &  &  &  & {\small{}$|H\rightarrow L+5\rangle$$-c.c.(0.2671)$}\tabularnewline
\hline 
{\small{}$VII_{x}$ } & {\small{}$10^{1}B\textrm{\ensuremath{_{3u}}}$ } & {\small{}$7.03$ } & {\small{}$0.510$ } & {\small{}$|H-2\rightarrow L;H\rightarrow L\rangle$$+c.c.(0.3680)$}\tabularnewline
 &  &  &  & {\small{}$|H-6\rightarrow L\rangle$$-c.c.(0.2715)$}\tabularnewline
\hline 
{\small{}$VIII_{x}$ } & {\small{}$13^{1}B\textrm{\ensuremath{_{3u}}}$ } & {\small{}$7.33$ } & {\small{}$0.402$ } & {\small{}$|H-1\rightarrow L+5\rangle$$-c.c.(0.4156)$}\tabularnewline
 &  &  &  & {\small{}$|H-3\rightarrow L+4\rangle$$-c.c.(0.2626)$}\tabularnewline
\hline 
\end{tabular}\label{tab:Linear absorption 16 atoms screened-1}
\end{table}

\begin{table}
\protect\caption{Excited states giving rise to the peaks in the linear absorption spectrum
of DQD-30, computed employing the MRSDCI approach along with standard
parameters in the PPP model Hamiltonian.}

\begin{tabular}{|c|c|c|c|c|}
\hline 
{\small{}Peak } & {\small{}State } & {\small{}E (eV) } & {\small{}Transition Dipole (Å) } & {\small{}Dominant Configurations}\tabularnewline
 &  &  &  & \tabularnewline
\hline 
{\small{}$I_{y}$ } & {\small{}$1^{1}B\textrm{\ensuremath{_{2u}}}$ } & {\small{}$2.31$ } & {\small{}$1.748$ } & {\small{}$|H\rightarrow L\rangle$$(0.8223)$}\tabularnewline
 &  &  &  & {\small{}$|H-1\rightarrow L+1\rangle$$(0.1632)$}\tabularnewline
\hline 
{\small{}$DF$} & {\small{}$1^{1}B\textrm{\ensuremath{_{3u}}}$ } & {\small{}$2.43$} & {\small{}$0$.000} & {\small{}$|H-2\rightarrow L\rangle$$-c.c.(0.4970)$}\tabularnewline
 &  &  &  & {\small{}$|H\rightarrow L+1;H\rightarrow L\rangle$$+c.c.(0.2365)$}\tabularnewline
\hline 
{\small{}$II_{x\&y}$ } & {\small{}$3^{1}B\textrm{\ensuremath{_{3u}}}$ } & {\small{}$3.83$ } & {\small{}$1.048$ } & {\small{}$|H-2\rightarrow L\rangle$$+c.c.(0.5432)$}\tabularnewline
 &  &  &  & {\small{}$|H-5\rightarrow L;H\rightarrow L\rangle+$$c.c.(0.1563)$}\tabularnewline
 & {\small{}$4^{1}B\textrm{\ensuremath{_{2u}}}$ } & {\small{}$3.85$ } & {\small{}$1.065$ } & {\small{}$|H-1\rightarrow L+1\rangle$$(0.4738)$}\tabularnewline
 &  &  &  & {\small{}$|H-4\rightarrow L\rangle$$-c.c.$ $(0.4659)$}\tabularnewline
 &  &  &  & {\small{}$|H-1\rightarrow L+1;H\rightarrow L;H\rightarrow L\rangle$$(0.1507)$}\tabularnewline
\hline 
{\small{}$III_{y}$ } & {\small{}$5^{1}B\textrm{\ensuremath{_{2u}}}$ } & {\small{}$4.20$ } & {\small{}$0.892$ } & {\small{}$|H-1\rightarrow L+1\rangle$$(0.5236)$}\tabularnewline
 &  &  &  & {\small{}$|H-4\rightarrow L\rangle$$-c.c.$ $(0.3332)$}\tabularnewline
\hline 
{\small{}$IV_{x}$ } & {\small{}$8^{1}B\textrm{\ensuremath{_{3u}}}$ } & {\small{}$4.74$ } & {\small{}$1.047$ } & {\small{}$|H-1\rightarrow L;H\rightarrow L\rangle$$-c.c.(0.3713)$}\tabularnewline
 &  &  &  & {\small{}$|H-1\rightarrow L+3\rangle$$+c.c.(0.2959)$}\tabularnewline
\hline 
{\small{}$V_{y}$ } & {\small{}$9^{1}B\textrm{\ensuremath{_{2u}}}$ } & {\small{}$5.10$ } & {\small{}$1.563$ } & {\small{}$|H-2\rightarrow L+2\rangle$$(0.5107)$}\tabularnewline
 &  &  &  & {\small{}$|H-7\rightarrow L\rangle$$-c.c.(0.2136)$}\tabularnewline
 &  &  &  & {\small{}$|H-3\rightarrow L;H\rightarrow L\rangle$$-c.c.(0.2088)$}\tabularnewline
 &  &  &  & {\small{}$|H-1\rightarrow L+1\rangle$$(0.2046)$}\tabularnewline
\hline 
{\small{}$VI_{x\&y}$ } & {\small{}$14^{1}B\textrm{\ensuremath{_{3u}}}$ } & {\small{}$5.66$ } & {\small{}$0.480$ } & {\small{}$|H-3\rightarrow L+1\rangle$$+c.c.(0.3604)$}\tabularnewline
 &  &  &  & {\small{}$|H\rightarrow L+1;H\rightarrow L\rangle$$-c.c.(0.2197)$}\tabularnewline
 &  &  &  & {\small{}$|H-8\rightarrow L\rangle+$$c.c.(0.1840)$}\tabularnewline
 & {\small{}$15^{1}B\textrm{\ensuremath{_{2u}}}$ } & {\small{}$5.70$ } & {\small{}$1.017$ } & {\small{}$|H-2\rightarrow L+2\rangle$$(0.3796)$}\tabularnewline
 &  &  &  & {\small{}$|H-3\rightarrow L+3\rangle$$(0.3475)$}\tabularnewline
 &  &  &  & {\small{}$|H-1\rightarrow L;H-2\rightarrow L\rangle$$-c.c.(0.2148)$}\tabularnewline
\hline 
{\small{}$VII_{y}$ } & {\small{}$20^{1}B\textrm{\ensuremath{_{2u}}}$ } & {\small{}$6.22$ } & {\small{}$1.113$ } & {\small{}$|H-4\rightarrow L+4\rangle$$(0.3458)$}\tabularnewline
 &  &  &  & {\small{}$|H-1\rightarrow L+1;H\rightarrow L;H\rightarrow L\rangle$$(0.2964)$}\tabularnewline
\hline 
\end{tabular}\label{tab:Linear absorption 30 atoms standard}
\end{table}

\begin{table}
\protect\caption{Excited states giving rise to the peaks in the linear absorption spectrum
of DQD-30, computed employing the MRSDCI approach along with screened
parameters in the PPP model Hamiltonian. }

\begin{tabular}{|c|c|c|c|c|}
\hline 
{\footnotesize{}Peak } & {\footnotesize{}State } & {\footnotesize{}E (eV) } & {\footnotesize{}Transition Dipole (Å) } & {\footnotesize{}Dominant Configurations}\tabularnewline
\hline 
{\footnotesize{}$I_{y}$ } & {\footnotesize{}$1^{1}B\textrm{\ensuremath{_{2u}}}$ } & {\footnotesize{}$2.08$ } & {\footnotesize{}$2.036$ } & {\footnotesize{}$|H\rightarrow L\rangle$$(0.8387)$}\tabularnewline
 &  &  &  & {\footnotesize{}$|H-1\rightarrow L+1\rangle$$(0.1389)$}\tabularnewline
\hline 
{\footnotesize{}$DF$} & {\footnotesize{}$1^{1}B\textrm{\ensuremath{_{3u}}}$ } & {\footnotesize{}$2.25$} & {\footnotesize{}$0.000$} & {\footnotesize{}$|H-2\rightarrow L\rangle$$+c.c.(0.4918)$}\tabularnewline
 &  &  &  & {\footnotesize{}$|H\rightarrow L+1;H\rightarrow L\rangle$$-c.c.(0.2691)$}\tabularnewline
\hline 
{\footnotesize{}$II_{x}$ } & {\footnotesize{}$3^{1}B\textrm{\ensuremath{_{3u}}}$ } & {\footnotesize{}$3.45$ } & {\footnotesize{}$1.443$ } & {\footnotesize{}$|H-2\rightarrow L\rangle-$$c.c.(0.5819)$}\tabularnewline
\hline 
{\footnotesize{}$III_{y}$ } & {\footnotesize{}$4^{1}B\textrm{\ensuremath{_{2u}}}$ } & {\footnotesize{}$3.71$ } & {\footnotesize{}$1.654$ } & {\footnotesize{}$|H-1\rightarrow L+1\rangle$$(0.7000)$}\tabularnewline
 &  &  &  & {\footnotesize{}$|H-4\rightarrow L\rangle+$$c.c.(0.2500)$}\tabularnewline
 &  &  &  & {\footnotesize{}$|H-1\rightarrow L+1;H\rightarrow L;H\rightarrow L\rangle$$(0.2466)$}\tabularnewline
\hline 
{\footnotesize{}$IV_{x}$ } & {\footnotesize{}$7^{1}B\textrm{\ensuremath{_{3u}}}$ } & {\footnotesize{}$4.23$ } & {\footnotesize{}$0.594$ } & {\footnotesize{}$|H-6\rightarrow L\rangle-$$c.c.(0.4901)$}\tabularnewline
 &  &  &  & {\footnotesize{}$|H-1\rightarrow L;H\rightarrow L\rangle+$$c.c.(0.2549)$}\tabularnewline
\hline 
{\footnotesize{}$V_{x\&y}$ } & {\footnotesize{}$11^{1}B\textrm{\ensuremath{_{3u}}}$ } & {\footnotesize{}$4.90$ } & {\footnotesize{}$1.119$ } & {\footnotesize{}$|H-3\rightarrow L+1\rangle+$$c.c.(0.5083)$}\tabularnewline
 &  &  &  & {\footnotesize{}$|H-5\rightarrow L;H\rightarrow L\rangle+$$c.c.(0.1442)$}\tabularnewline
 &  &  &  & {\footnotesize{}$|H-7\rightarrow L+1;H\rightarrow L\rangle$$-$$c.c.(0.1350)$}\tabularnewline
 & {\footnotesize{}$10^{1}B\textrm{\ensuremath{_{2u}}}$ } & {\footnotesize{}$4.83$ } & {\footnotesize{}$1.221$ } & {\footnotesize{}$|H-2\rightarrow L+2\rangle$$(0.5634)$}\tabularnewline
 &  &  &  & {\footnotesize{}$|H-2\rightarrow L+1;H\rightarrow L\rangle$$c.c.(0.2359)$}\tabularnewline
 &  &  &  & {\footnotesize{}$|H-1\rightarrow L+5\rangle$$+c.c.(0.2065)$}\tabularnewline
\hline 
{\footnotesize{}$VI_{y}$ } & {\footnotesize{}$15^{1}B\textrm{\ensuremath{_{2u}}}$ } & {\footnotesize{}$5.32$ } & {\footnotesize{}$0.632$ } & {\footnotesize{}$|H\rightarrow L+3;H\rightarrow L\rangle+$$c.c.(0.3245)$}\tabularnewline
 &  &  &  & {\footnotesize{}$|H-7\rightarrow L\rangle-$$c.c.(0.3079)$}\tabularnewline
 &  &  &  & {\footnotesize{}$|H-1\rightarrow L;H-2\rightarrow L\rangle-$$c.c.(0.2057)$}\tabularnewline
\hline 
{\footnotesize{}$VII_{x}$ } & {\footnotesize{}$20^{1}B\textrm{\ensuremath{_{3u}}}$ } & {\footnotesize{}$5.69$ } & {\footnotesize{}$0.431$ } & {\footnotesize{}$|H-2\rightarrow L+4\rangle-$$c.c.(0.4019)$}\tabularnewline
 &  &  &  & {\footnotesize{}$|H-1\rightarrow L+1;H\rightarrow L+1\rangle+$$c.c.(0.2542)$}\tabularnewline
 &  &  &  & {\footnotesize{}$|H-1\rightarrow L;H-4\rightarrow L\rangle+$$c.c.(0.2371)$}\tabularnewline
\hline 
{\footnotesize{}$VIII_{x}$ } & {\footnotesize{}$23^{1}B\textrm{\ensuremath{_{3u}}}$ } & {\footnotesize{}$5.97$ } & {\footnotesize{}$0.498$ } & {\footnotesize{}$|H-7\rightarrow L+2\rangle+$$c.c.(0.3404)$}\tabularnewline
 &  &  &  & {\footnotesize{}$|H\rightarrow L+2;H-3\rightarrow L\rangle$$c.c.(0.3106)$}\tabularnewline
\hline 
{\footnotesize{}$IX_{x}$ } & {\footnotesize{}$29^{1}B\textrm{\ensuremath{_{3u}}}$ } & {\footnotesize{}$6.28$ } & {\footnotesize{}$1.063$ } & {\footnotesize{}$|H-3\rightarrow L+5\rangle+$$c.c.(0.3091)$}\tabularnewline
 &  &  &  & {\footnotesize{}$|H-6\rightarrow L+4\rangle-$$c.c.(0.2390)$}\tabularnewline
\hline 
\end{tabular}\label{tab:Linear absorption 30 atoms screened}
\end{table}

\begin{table}
\protect\caption{Excited states giving rise to the peaks $I_{y}$ to $VII_{y}$ in
the linear absorption spectrum of DQD-48, computed employing the MRSDCI
approach along with standard parameters in the PPP model Hamiltonian.}

\begin{tabular}{|c|c|c|c|c|}
\hline 
{\small{}Peak } & {\small{}State } & {\small{}E (eV) } & {\small{}Transition Dipole (Å) } & {\small{}Dominant Configurations}\tabularnewline
 &  &  &  & \tabularnewline
\hline 
{\small{}$DF$} & {\small{}$1^{1}B\textrm{\ensuremath{_{3u}}}$ } & {\small{}$1.52$} & {\small{}$0$.000} & {\small{}$|H\rightarrow L+1;H\rightarrow L\rangle-$$c.c.(0.4873)$}\tabularnewline
 &  &  &  & {\small{}$|H-4\rightarrow L\rangle+$$c.c.(0.2313)$}\tabularnewline
\hline 
{\small{}$I_{y}$ } & {\small{}$1^{1}B\textrm{\ensuremath{_{2u}}}$ } & {\small{}$1.61$ } & {\small{}$1.402$ } & {\small{}$|H\rightarrow L\rangle$$(0.7467)$}\tabularnewline
 &  &  &  & {\small{}$|H-1\rightarrow L+1\rangle$$(0.3061)$}\tabularnewline
\hline 
{\small{}$II_{y}$ } & {\small{}$3^{1}B\textrm{\ensuremath{_{2u}}}$ } & {\small{}$2.57$ } & {\small{}$2.180$ } & {\small{}$|H-1\rightarrow L+1\rangle$$(0.5160)$}\tabularnewline
 &  &  &  & {\small{}$|H-1\rightarrow L+1;H\rightarrow L;H\rightarrow L\rangle$$(0.4413)$}\tabularnewline
 &  &  &  & {\small{}$|H-2\rightarrow L\rangle$$+c.c.(0.2739)$}\tabularnewline
\hline 
{\small{}$III_{x}$ } & {\small{}$3^{1}B\textrm{\ensuremath{_{3u}}}$ } & {\small{}$2.76$ } & {\small{}$1.043$ } & {\small{}$|H\rightarrow L+1;H\rightarrow L\rangle+$$c.c.(0.4903)$}\tabularnewline
 &  &  &  & {\small{}$|H\rightarrow L+5\rangle-$$c.c.(0.2202)$}\tabularnewline
 &  &  &  & {\small{}$|H\rightarrow L+1;H-1\rightarrow L+1\rangle+$$c.c.(0.1449)$}\tabularnewline
\hline 
{\small{}$IV_{y}$ } & {\small{}$5^{1}B\textrm{\ensuremath{_{2u}}}$ } & {\small{}$2.88$ } & {\small{}$1.067$ } & {\small{}$|H-2\rightarrow L\rangle+$$c.c.(0.4765)$}\tabularnewline
 &  &  &  & {\small{}$|H-1\rightarrow L+1\rangle$$(0.2689)$}\tabularnewline
 &  &  &  & {\small{}$|H-1\rightarrow L+1;H\rightarrow L;H\rightarrow L\rangle$$(0.2549)$}\tabularnewline
\hline 
{\small{}$V_{x}$ } & {\small{}$6^{1}B\textrm{\ensuremath{_{3u}}}$ } & {\small{}$3.38$ } & {\small{}$0.933$ } & {\small{}$|H-4\rightarrow L\rangle-$$c.c.(0.4874)$}\tabularnewline
 &  &  &  & {\small{}$|H-6\rightarrow L;H\rightarrow L\rangle+$$c.c.(0.2237)$}\tabularnewline
 &  &  &  & {\small{}$|H-3\rightarrow L+1\rangle-$$c.c.(0.1679)$}\tabularnewline
\hline 
{\small{}$VI_{y}$ } & {\small{}$6^{1}B\textrm{\ensuremath{_{2u}}}$ } & {\small{}$3.66$ } & {\small{}$1.727$ } & {\small{}$|H-3\rightarrow L;H\rightarrow L\rangle$$(0.3471)$}\tabularnewline
 &  &  &  & {\small{}$|H-1\rightarrow L+1\rangle$$(0.2908)$}\tabularnewline
\hline 
{\small{}$VII_{y}$ } & {\small{}$10^{1}B\textrm{\ensuremath{_{2u}}}$ } & {\small{}$4.16$ } & {\small{}$0.939$ } & {\small{}$|H-2\rightarrow L+2\rangle$$(0.3609)$}\tabularnewline
 &  &  &  & {\small{}$|H-1\rightarrow L+7\rangle+$$c.c.(0.3331)$}\tabularnewline
\hline 
\end{tabular}\label{tab:Linear absorption 48 atoms standard-1}
\end{table}

\begin{table}
\protect\caption{Excited states giving rise to the peaks $VIII_{y}$ to $XII_{y}$
in the linear absorption spectrum of DQD-48, computed employing the
MRSDCI approach along with standard parameters in the PPP model Hamiltonian.}

\begin{tabular}{|c|c|c|c|c|}
\hline 
{\small{}Peak } & {\small{}State } & {\small{}E (eV) } & {\small{}Transition Dipole (Å) } & {\small{}Dominant Configurations}\tabularnewline
 &  &  &  & \tabularnewline
\hline 
{\small{}$VIII_{y}$ } & {\small{}$13^{1}B\textrm{\ensuremath{_{2u}}}$ } & {\small{}$4.37$ } & {\small{}$0.596$ } & {\small{}$|H-4\rightarrow L+1;H\rightarrow L\rangle$$c.c.(0.3129)$}\tabularnewline
 &  &  &  & {\small{}$|H-4\rightarrow L+4\rangle$$(0.2685)$}\tabularnewline
 &  &  &  & {\small{}$|H-1\rightarrow L;H-4\rightarrow L\rangle-$$c.c.(0.2191)$}\tabularnewline
\hline 
{\small{}$IX_{x\&y}$ } & {\small{}$13^{1}B\textrm{\ensuremath{_{3u}}}$ } & {\small{}$4.59$ } & {\small{}$0.839$ } & {\small{}$|H-3\rightarrow L+1\rangle-$$c.c.(0.4035)$}\tabularnewline
 &  &  &  & {\small{}$|H-8\rightarrow L;H\rightarrow L+1\rangle$$c.c.(0.1534)$}\tabularnewline
 &  &  &  & {\small{}$|H\rightarrow L+5\rangle-$$c.c.(0.1371)$}\tabularnewline
\hline 
 & {\small{}$18^{1}B\textrm{\ensuremath{_{2u}}}$ } & {\small{}$4.65$ } & {\small{}$0.231$ } & {\small{}$|H\rightarrow L+4;H\rightarrow L+1\rangle-$$c.c.(0.2687)$}\tabularnewline
 &  &  &  & {\small{}$|H-6\rightarrow L+1\rangle+$$c.c.(0.2665)$}\tabularnewline
 &  &  &  & {\small{}$|H\rightarrow L+9;H\rightarrow L\rangle+$$c.c.(0.2156)$}\tabularnewline
\hline 
{\small{}$X_{y}$ } & {\small{}$20^{1}B\textrm{\ensuremath{_{2u}}}$ } & {\small{}$4.80$ } & {\small{}$0.347$ } & {\small{}$|H-4\rightarrow L+1;H\rightarrow L\rangle$$c.c.(0.2900)$}\tabularnewline
 &  &  &  & {\small{}$|H-1\rightarrow L+5;H\rightarrow L\rangle$$c.c.(0.1942)$}\tabularnewline
 &  &  &  & {\small{}$|H-2\rightarrow L+2\rangle$$(0.1894)$}\tabularnewline
\hline 
{\small{}$XI_{x\&y}$ } & {\small{}$20^{1}B\textrm{\ensuremath{_{3u}}}$ } & {\small{}$4.96$ } & {\small{}$0.237$ } & {\small{}$|H-9\rightarrow L+1\rangle+$$c.c.(0.4364)$}\tabularnewline
 &  &  &  & {\small{}$|H-8\rightarrow L;H\rightarrow L+1\rangle$$c.c.(0.1441)$}\tabularnewline
 &  &  &  & {\small{}$|H-10\rightarrow L;H\rightarrow L\rangle$$-c.c.(0.1418)$}\tabularnewline
\hline 
 & {\small{}$23^{1}B\textrm{\ensuremath{_{2u}}}$ } & {\small{}$4.98$ } & {\small{}$0.634$ } & {\small{}$|H-2\rightarrow L+2\rangle$$(0.2775)$}\tabularnewline
 &  &  &  & {\small{}$|H-1\rightarrow L+5;H\rightarrow L\rangle$$c.c.(0.2171)$}\tabularnewline
 &  &  &  & {\small{}$|H-4\rightarrow L+1;H\rightarrow L\rangle$$c.c.(0.2162)$}\tabularnewline
\hline 
{\small{}$XII_{y}$ } & {\small{}$33{}^{1}B\textrm{\ensuremath{_{2u}}}$ } & {\small{}$5.54$ } & {\small{}$1.183$ } & {\small{}$|H-4\rightarrow L+4\rangle$$(0.2954)$}\tabularnewline
 &  &  &  & {\small{}$|H-13\rightarrow L\rangle+$$c.c.(0.2185)$}\tabularnewline
 &  &  &  & {\small{}$|H-1\rightarrow L+1;H\rightarrow L+3\rangle$$c.c.(0.2098)$}\tabularnewline
\hline 
\end{tabular}\label{tab:Linear absorption 48 atoms standard}
\end{table}

\begin{table}
\protect\caption{Excited states giving rise to the peaks in the linear absorption spectrum
of DQD-48, computed employing the MRSDCI approach along with screened
parameters in the PPP model Hamiltonian.}

\begin{tabular}{|c|c|c|c|c|}
\hline 
{\small{}Peak } & {\small{}State } & {\small{}E (eV) } & {\small{}Transition } & {\small{}Dominant Configurations}\tabularnewline
 &  &  & {\small{}Dipole (Å)} & \tabularnewline
\hline 
{\small{}$DF$} & {\small{}$1^{1}B\textrm{\ensuremath{_{3u}}}$ } & {\small{}$1.38$} & {\small{}$0$.000} & {\small{}$|H\rightarrow L+1;H\rightarrow L\rangle-$$c.c.(0.5106)$}\tabularnewline
 &  &  &  & {\small{}$|H\rightarrow L+3\rangle-c.c.$$(0.2285)$}\tabularnewline
\hline 
{\small{}$I_{y}$ } & {\small{}$1^{1}B\textrm{\ensuremath{_{2u}}}$ } & {\small{}$1.40$ } & {\small{}$1.770$ } & {\small{}$|H\rightarrow L\rangle$$(0.7890)$}\tabularnewline
 &  &  &  & {\small{}$|H-1\rightarrow L+1\rangle$$(0.2338)$}\tabularnewline
\hline 
{\small{}$II_{y}$ } & {\small{}$2^{1}B\textrm{\ensuremath{_{2u}}}$ } & {\small{}$2.19$ } & {\small{}$2.910$ } & {\small{}$|H-1\rightarrow L+1\rangle$$(0.6593)$}\tabularnewline
 &  &  &  & {\small{}$|H-1\rightarrow L+1;H\rightarrow L;H\rightarrow L\rangle$$(0.4735)$}\tabularnewline
\hline 
{\small{}$III_{x}$ } & {\small{}$6^{1}B\textrm{\ensuremath{_{3u}}}$ } & {\small{}$3.05$ } & {\small{}$0.984$ } & {\small{}$|H-3\rightarrow L\rangle+$$c.c.(0.4918)$}\tabularnewline
 &  &  &  & {\small{}$|H-1\rightarrow L+1;H\rightarrow L+1\rangle+$$c.c.(0.1791)$}\tabularnewline
\hline 
{\small{}$IV_{x}$ } & {\small{}$8^{1}B\textrm{\ensuremath{_{3u}}}$ } & {\small{}$3.32$ } & {\small{}$1.135$ } & {\small{}$|H\rightarrow L+1;H-2\rightarrow L\rangle$$c.c.(0.3569)$}\tabularnewline
 &  &  &  & {\small{}$|H-1\rightarrow L+1;H\rightarrow L+1\rangle+$$c.c.(0.3481)$}\tabularnewline
\hline 
{\small{}$V_{y}$ } & {\small{}$9^{1}B\textrm{\ensuremath{_{2u}}}$ } & {\small{}$3.61$ } & {\small{}$1.313$ } & {\small{}$|H\rightarrow L+4;H\rightarrow L\rangle+$$c.c.(0.3971)$}\tabularnewline
 &  &  &  & {\small{}$|H-9\rightarrow L\rangle+$$c.c.(0.2276)$}\tabularnewline
\hline 
{\small{}$VI_{x}$ } & {\small{}$16^{1}B\textrm{\ensuremath{_{3u}}}$ } & {\small{}$3.96$ } & {\small{}$1.163$ } & {\small{}$|H-4\rightarrow L+1\rangle+$$c.c.(0.3369)$}\tabularnewline
 &  &  &  & {\small{}$|H\rightarrow L+1;H-2\rightarrow L\rangle$$c.c.(0.2805)$}\tabularnewline
\hline 
{\small{}$VII_{y}$ } & {\small{}$24^{1}B\textrm{\ensuremath{_{2u}}}$ } & {\small{}$4.38$ } & {\small{}$0.661$ } & {\small{}$|H-1\rightarrow L+1;H-1\rightarrow L+1;H\rightarrow L\rangle$$(0.2485)$}\tabularnewline
 &  &  &  & {\small{}$|H-2\rightarrow L+2\rangle$$(0.2256)$}\tabularnewline
 &  &  &  & {\small{}$|H\rightarrow L+1;H\rightarrow L+1;H-2\rightarrow L\rangle+$$c.c.(0.1943)$}\tabularnewline
\hline 
{\small{}$VIII_{y}$ } & {\small{}$33^{1}B\textrm{\ensuremath{_{2u}}}$ } & {\small{}$4.78$ } & {\small{}$0.520$ } & {\small{}$|H-1\rightarrow L+1;H\rightarrow L+4\rangle$$c.c.(0.3444)$}\tabularnewline
 &  &  &  & {\small{}$|H-1\rightarrow L+5;H\rightarrow L\rangle$$c.c.(0.1999)$}\tabularnewline
 &  &  &  & {\small{}$|H-1\rightarrow L+10\rangle-$$c.c.(0.1855)$}\tabularnewline
\hline 
\end{tabular}\label{tab:Linear absorption 48 atoms screened}
\end{table}

\newpage{}

\bibliographystyle{apsrev4-1}
\bibliography{Graphene}

\end{document}